\documentclass[twocolumn]{aastex631}

\usepackage{amsmath}
\usepackage{multirow}
\usepackage{CJK}
\usepackage{natbib}

\shorttitle{Infall in AB~Aur}
\shortauthors{Calcino et al.}

\graphicspath{{./}{figures/}}
\begin{document}
\begin{CJK*}{UTF8}{gbsn}

% \title[Infall Induced Spiral Arms in the Disk Kinematics Around AB~Aur]{Infall Induced Wiggles in the Disk Kinematics of AB~Aur}

\title[Infall in AB~Aur]{Infall Explains the Disk Kinematics of AB Aur Without Gravitational Instability}

\author[0000-0001-7764-3627]{Josh Calcino}
\altaffiliation{jcalcino@tsinghua.edu.cn}
\affiliation{Department of Astronomy, Tsinghua University, 30 Shuangqing Rd, 100084 Beijing, China}
\author[0000-0002-4716-4235]{Daniel J. Price}
\affiliation{School of Physics \& Astronomy, Monash University VIC 3800}
\affiliation{Univ. Grenoble Alpes, CNRS, IPAG, 38000 Grenoble, France}
\author[0000-0003-4672-8411]{Chris W. Ormel}
\affiliation{Department of Astronomy, Tsinghua University, 30 Shuangqing Rd, 100084 Beijing, China}

\defcitealias{bollati2021}{B21}

\begin{abstract}
Late-stage infall onto protoplanetary disks can produce large scale spiral arms. In this paper we used 3D smoothed particle hydrodynamics and radiative transfer simulations to study the kinematic perturbations induced in disks by infalling material. We found that deviations from Keplerian rotation are predominantly in the radial and vertical velocity components, spatially correlated with spiral arms in the gas surface density. The infall produces observable wiggles in the channel maps, analogous to those produce by the gravitational instability (GI), along with large-scale arcs and filaments. GI induced spiral arms produce radial velocity perturbations that point towards the center of the spiral arm owing to their higher self-gravity. We found a similar signature from infall-induced spiral arms, despite not including self-gravity in our simulation. Our study suggests that recent evidence of GI in the kinematics of the disk around AB~Aur may instead be due to the observed infall, without the need for invoking GI. We further show that a unified model invoking infall onto a central binary can explain the CO morphology and kinematics, scattered light spirals, and mm-continuum emission in AB Aur.
%Our study suggests that the recent evidence of gravitational instability in the disk around AB~Aur could be entirely explained by the observed infall. 
\end{abstract}


\section{Introduction} \label{sec:intro}

Recent scattered light and molecular line observations have revealed that a large fraction of Class II protoplanetary disks interact with their environment via `late-stage' accretion \citep{huang2020, huang2021, Mesa2022, garufi2022, benisty2023, garufi2024, gupta2024}. 
Typically characterized as filamentary-like structures infalling on ballistic trajectories, late-stage accretion changes the mass and angular momentum evolution of the disk \citep{winter2024,pelkonen2025}. 
The exact effect of late-stage accretion on protoplanetary disk substructures are not well understood. Simulations have shown that abundant spiral structures \citep{dullemond2019, kuffmeier2021, calcino2025, huhn2026}, misaligned disks \citep{kuffmeier2021}, vortices \citep{bae2015}, and dust gaps \citep{Kuznetsova2022} can be formed. 

All of the substructures infall can produce in a protoplanetary disk can also be produced by other mechanisms. For example, spiral arms and dust gaps can be produced by planets \citep[e.g.][]{dipierro2015,zhang2018}, while spiral arms can be produced by the gravitational instability (GI) \citep{lodato2005}. Determining which mechanism generates a particular substructure is challenging using the disk morphology alone. Recently, \cite{calcino2025} showed that streamers can produce multiple spiral arms that move with different pattern speeds, ranging from stationary to Super-Keplerian, which can distinguish infall from GI. However measuring the pattern speeds of spiral arms in the outer several hundred au of the disk requires high-spatial resolution observations with temporal baselines of roughly a decade \citep[e.g.][]{ren2018}. Such observations are lacking for the majority of Class II disks.

Another method is to study the disk kinematics \citep[e.g.][]{pinte2023}. Many studies have investigated the detectable kinematics that companions, of either planetary \citep{perez2018, pinte2018, calcino2020, bollati2021} or stellar mass \citep{price2018,calcino2019, norfolk2022, calcino2023, calcino2024}, can have in the disk. Spiral arms launched by planets produce predominantly radial velocity perturbations which are divergent \citep{goodman2001, rafikov2002, bollati2021, calcino2022}. That is, the radial velocity residuals have motion that is moving away from the center of the spiral arm. In contrast, owing to the stronger self-gravity inside the spirals than outside, spiral arms generated by GI have radial velocities that are radially convergent \citep{hall2020, Longarini2021}. This signature was recently found in the disk around AB~Aur \citep{speedie2024}, and suggests the spiral arms may be generated by GI. However AB~Aur is known to be experiencing infall \citep{speedie2025}, and the direct effect of this infall on the disk structure and kinematics is not understood.

It has been argued by previous authors \citep{fukagawa2004,cadman2021,speedie2024} that the disk of AB Aur is gravitationally unstable in order to explain the large-scale spiral arms. However, for GI the disk must be unstable both globally $M_{\rm disk}/M_{*} \gtrsim H/R$ and locally $Q < 1$ according to the \citet{toomre1964} criterion on the basis of the large scale spiral arms. Disk mass estimates in AB Aur are uncertain, but the present-day measurements point to numbers below the GI threshold. \citet{tang2012-abaur-envelope} inferred a total gas mass of $\sim 10^{-2} \,{\rm M}_\odot$ from the dust continuum emission which would not satisfy either criteria. If the dust is optically thick this could be a lower limit, but they also found a surprisingly low gas mass in the spiral arms ($10^{-7}$--$10^{-5} \,{\rm M}_\odot$) indicating a low mass, gravitationally stable disk (as also found by \citealt{pietu2005}). Similarly \citet{woitke2019} inferred a gas mass of only $0.019 {\rm M}_{\odot}$ from thermochemical model fitting to the molecular line emission, also close to their estimate of $0.022 {\rm M}_{\odot}$ from their pure SED fit, implying $M_{\rm disk}/M_* = 0.88\%$ (their table 4). Finally \citet{kama2020} used the HD 1-0 line to get an upper limit on the disc gas mass of $\leq 0.1 {\rm M}_\odot$ and hence argued that the disk around AB~Aur cannot be gravitationally unstable, since GI would require at least $M_{\rm disk} \gtrsim 0.3 {\rm M}_\odot$ \citep{cadman2021}.

In contrast, the main evidence for GI in AB Aur is the disturbed kinematics \citep{speedie2024}, in particular the large scale `GI Wiggle' seen in channel maps, has been identified by \citet{hall2020}  as the main kinematic signature of GI (see also \citealt{Longarini2021,terry2022}).

In this Letter, we test the hypothesis that the large scale kinematic perturbations and GI Wiggle seen in AB Aur can be explained by the observed infall \citep{speedie2025} without requiring gravitational instability. In addition we show that infall onto a circumbinary disc can reproduce the central dust cavity and the complex spiral morphology in AB~Aur.

\section{Methods}

\subsection{Hydrodynamical Simulations}

We used the smoothed particle hydrodynamics (SPH; \citealt{monaghan1992,price2012}) code {\sc{phantom}} \citep{phantom2018} to simulate the interaction of infalling streamers. Our simulation setup is identical to that of \cite{calcino2025} except that the physical units have been rescaled to better match the stellar mass and disk size in AB~Aur. For simplicity, we ignore magnetic fields \citep[][]{unno2022}, while the disk self-gravity is negligible in our model because of the low assumed disk mass. For the initial set of simulations we performed gas-only simulations, assuming perfectly coupled dust, but performed an additional simulation with dust decoupling that is discussed in Section \ref{sec:ab_aur_comp}. The simulations begin from an isolated, locally isothermal disk with no infalling material which is evolved for approximately 8 outer orbits of the disk.  For the gas-only simulations we distributed $2\times10^{6}$ SPH particles initially between $R_\mathrm{in} = 9\,\mathrm{au}$ and $R_\mathrm{out} = 630\,\mathrm{au}$, following a surface density profile \(\Sigma(R) \propto R^{-1}\). The central star is represented by a \(2.5\,\mathrm{M}_\odot\) sink particle with an accretion radius of \(r_\mathrm{acc} = 9\,\mathrm{au}\) \citep{bate1995}. We adopt a locally isothermal equation of state with a sound speed profile \(c_\mathrm{s}(R) \propto R^{-0.25}\), corresponding to temperatures \(T(R) \propto R^{-0.5}\) and a disk aspect ratio \(H/R = 0.11\) at \(R = 630\,\mathrm{au}\). Since we assume the disc self-gravity is negligible and use a locally isothermal equation of state, we are free to rescale the gas mass of our gas-only simulations when performing radiative transfer.
Disk viscosity is included via an SPH viscosity parameter \(\alpha_\mathrm{AV} = 0.1\), which corresponds to an effective Shakura--Sunyaev viscosity \(\alpha_\mathrm{SS} \approx 2.5\times10^{-3}\) \citep{lodato2010}. 

\subsection{Initializing the Infall}
\begin{table}
\centering
\caption{Parameters of the infalling streamers in each simulation. Run 1 contains one streamer, while Run 2 includes two streamers. The streamers are characterized by their mass, ellipse parameters \(a\) and \(b\), pericenter radius \(r_\mathrm{peri}\), initial distance \(r_\mathrm{init}\), and inclination angles \((i_\mathrm{inf}, \phi_\mathrm{inf})\).}
\label{tab:streamer_params}
\begin{tabular}{lcccccc}
\hline
\textbf{Run} & \textbf{Str.} & \textbf{Mass} & \(a\) & \(b\) & \(r_\mathrm{peri}\) & \((i_\mathrm{inf}, \phi_\mathrm{inf})\) \\
             &               & [M$_\textrm{Disk}$] & [au] & [au] & [au] & [deg] \\
\hline
\multirow{1}{*}{Run 1} & 1 & 10\% & 360 & 36 & 270 & (50, 0) \\
\multirow{2}{*}{Run 2} & 1 & 5\% & 1080 & 36 & 270 & (30, 0) \\
                       & 2 & 5\% & 1080 & 36 & 270 & (50, 30) \\
\hline
\end{tabular}
\end{table}

We performed two simulations which differ in the initial conditions of the streamer material. Both simulations are used as input models for radiative transfer, with the first simulation acting as a test-bed to understand the gas kinematics of streamer-disk interactions, while the second is post-processed to generate synthetic observations for comparison with AB~Aur. As in \cite{calcino2025}, we model our streamers as ellipses with initial semi-minor axis $b$ and a semi-major axis $a$ on an free-falling parabolic orbit. The elliptical streamer is initialized with its center at radius $r_\textrm{init}$ along a parabola with a periastron distance of $r_\textrm{peri}$. 

In the first simulation, since we are primarily concerned with the kinematics of streamer-disk interactions, we include only one inclined streamer. The dimensions of the streamer are $a=360$ au and $b=36$ au, initialized at a radius $r_\textrm{init} = 1080$ au, periastron passage distance of $r_\textrm{peri} = 270$ au, and a mass $M_\textrm{in} = 10\% M_\textrm{disk}$. The streamer is initialized with an inclination above the disk midplane of $50^{\circ}$.

In AB~Aur, two distinct streamers are observed interacting and merging with the disk with a bright spot in SO emission \citep{speedie2025}. Thus, in the second simulation we initialize two streamers with $a=1080$ au and $b=36$ au, with a mass $M_\textrm{in} = 5\% M_\textrm{disk}$, with the first and second streamers inclined $30^\circ$ and $50^\circ$ degrees above the midplane, respectively. The second streamer is rotated $30^\circ$ in a clockwise direction away from the first. 

% $M_\textrm{in} = 1\times10^{-4}$ M$_\odot$ (5\% of the disk mass each), while the first and second streamers inclined $30^\circ$ and $50^\circ$ degrees above the midplane, respectively. The second streamer is rotated $30^\circ$ in a clockwise direction away from the first. 

The infall rate to the disk can be estimated by assuming that all material accretes onto the disk at the pericenter of the parabolic orbit using Barker's equation. Doing-so yields a time-difference between the tip and tail of the streamer reaching pericenter as $\sim 1500$ years and $\sim 5100$ years for the first and second simulations, respectively. The precise infall rate on the disk depends on the assumed gas mass. Using $M_\textrm{gas} = 0.05\,{\rm M}_\odot$ this gives a time-averaged $\dot{M}_\textrm{inf}\sim3.1\times10^{-6} \,{\rm M}_\odot\, {\rm yr}^{-1}$ and $\dot{M}_\textrm{inf}\sim8.9\times10^{-7} \,{\rm M}_\odot\, {\rm yr}^{-1}$ respectively. Although these infall rates are higher than the estimated stellar accretion rates of AB~Aur of $\dot{M}_\star\sim10^{-7} \,{\rm M}_\odot\, {\rm yr}^{-1}$ \citep{Garcia-Lopez+2006,brittain2007,donehew2011,salyk2013}, material landing on the disk must first be processed by the disk before accreting onto the central star. Both simulations \citep{offner2012, kuffmeier2018b} and observations \citep{valdiviamena2022, flores2023} suggest infall rates onto the disk can be an order of magnitude larger than the accretion rate onto the central star. This is also seen in our simulations, where the accretion rate onto the central sink peaks at $\sim 2\times 10^{-7}\,{\rm M}_\odot\, {\rm yr}^{-1}$. In addition, not all streamer material immediately merges with the disk during their interaction. As in \cite{calcino2025}, some portion of the streamer material is close to being unbound and is launched into a wide orbit, taking longer to accrete than the timescale estimated above. Therefore our estimated accretion rates should be taken as an upper limit. 

% which is comparable with the stellar accretion rate estimate of AB Aur of $\dot{M}_\star\sim10^{-7} \,{\rm M}_\odot\, {\rm yr}^{-1}$ \citep{Garcia-Lopez+2006,brittain2007,donehew2011,salyk2013}. 
\subsection{Radiative Transfer and Mock Observations}

Our SPH simulations were used to generate a Voronoi mesh for input in the Monte Carlo radiative transfer code {\sc{mcfost}} \citep{pinte2006, pinte2009}. For the gas-only simulations, we assumed dust particles perfectly coupled to the gas following a power-law grain size distribution $dn/ds \propto s^{-3.5}$ for $0.03\mu$m $\leq s \leq 10\mu$m. We scale the gas-to-dust ratio such that the mass of grains between this limited range is the same as a having a power-law grain size distribution $dn/ds \propto s^{-3.5}$ for $0.03\mu$m $\leq s \leq 1$mm with a gas-to-dust ratio of 100. We use a relatively small upper limit on the grain size since the opacity on the surface of the disc is dominated by small grains. We set the total gas mass inside the domain to $M_\textrm{gas} = 0.05\ M_\odot$. For our dusty simulation in Section \ref{sec:ab_aur_comp} we used the full dust grain size distribution $dn/ds \propto s^{-3.5}$ for $0.03\mu$m $\leq s \leq 1$mm, and do not rescale the gas mass.

For the radiative transfer we assumed spherical and homogeneous grains composed of astronomical silicate \citep{weingartner2001}. The central star was set with an effective temperature $T_\textrm{eff} = 9970$ K and a radius $R_\star = 2.5 $ R$_\odot$. We assumed an accretion rate of $1.3\times10^{-7}\ $M$_\odot$ onto the central star.

The temperature profile and specific intensities were generated using $10^8$ photon packets. Images were produced by ray-tracing the computed source function. Our CO isotopologue observations were generated assuming $T_\textrm{gas} = T_\textrm{dust}$ with all molecules in Local Thermodynamical Equilibrium (LTE). We assumed an initially constant $^{13}$CO abundance similar to the ISM with ratio $^{13}$CO/H$_2 = 2.0\times 10^{-6}$ \citep{wilson1994, milam2005}. This abundance was then altered due to CO freeze out ($T = 20$ K, depletion factor of $10^{-4}$), photo-desorption, and photo-dissociation, following Appendix~B of \cite{pinte2018b}.

When making mock observations of Run 1, we set the disk inclination to 30$^{\circ}$, the disk Position Angle (PA) to $270^{\circ}$, and distance to 155.9 pc. We convolved the resulting $^{13}$CO (2-1) mock observations with a 0.1$^{\arcsec}$ beam but did not add any noise. For our match to the AB~Aur system, we inclined Run 2 by $23.2^{\circ}$ and set the disk PA to $236.7^{\circ}$. We convolved the observations with a $237 \times 175$ mas beam and add random noise with an rms of 2.0 mJy/beam. These conditions match the Briggs robust 0.5 weighted image cubes from \citep{speedie2024}. Moment maps were created with {\sc bettermoments} \citep{bettermoments2018}. 

% fiducial model is inclined 30 degrees, disk PA is 270 Distance 159 pc. Mgas is forced in the radiative transfer to be equal to .. Cannot do this, need to rerun RT models. 
% ab~aur compare: inc 23.2, PA 236.7, distance 155.9, as in Speedie et al. 

% synth obs: 
% fiducial: assume no noise, beam of 0.1 arcsec for ALMA obs, no beam smearing for sphere since scale it's so small anyway 
% ab~aur compare: use 237 x 175 mas beam and noise level of 2.0 mJy, same as the robust 0.5 observation of AB~aur in \citep{speedie2025}

\section{Results}\label{sec:res}

% \begin{enumerate}
%     \item surface density and velocity plots 
%     \item radiative transfer model showing channels, for full model, keplerian, only radial, only azimuthal?
%     \item velocity maps with Jess's filter
%     \item direct comparison with AB Aur
%     \item the wiggle along the minor axis
% \end{enumerate}

\begin{figure*}
\centering
\includegraphics[width=\textwidth]{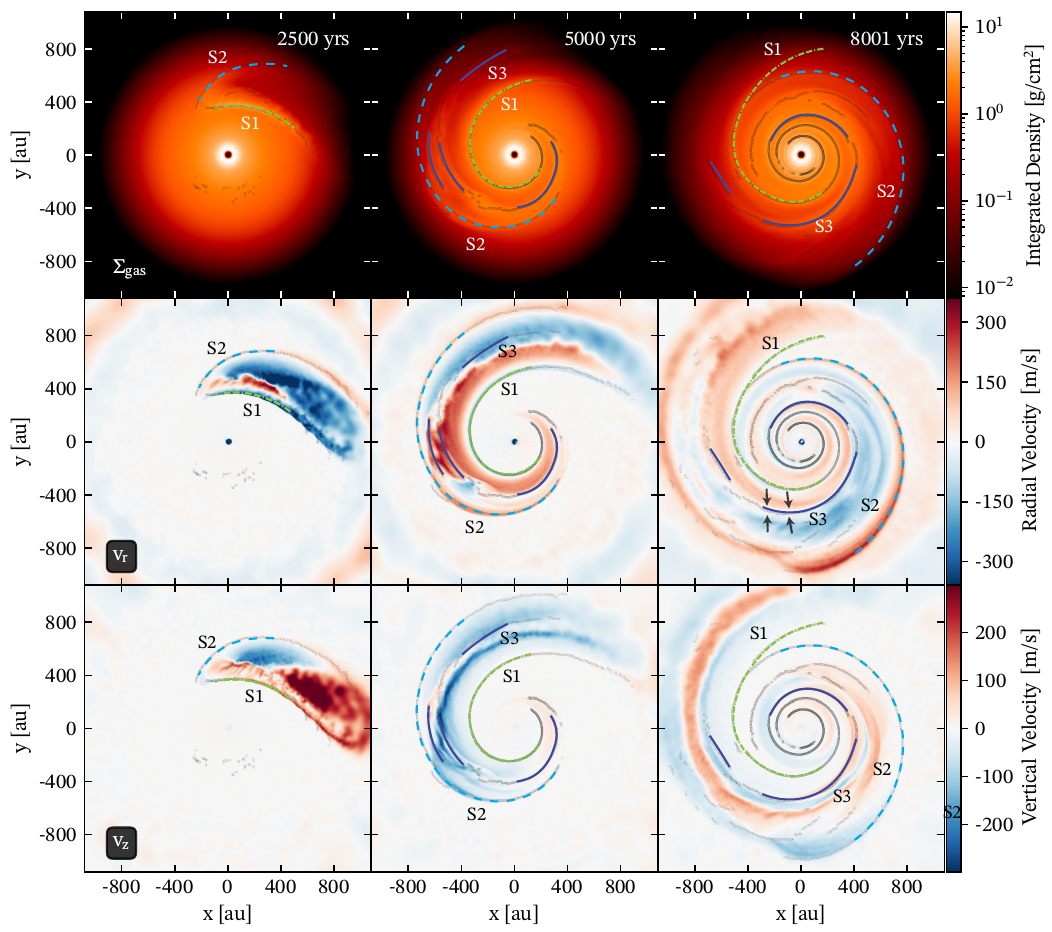}
\caption{Vertically integrated surface density (top row), midplane radial velocities (middle row) and midplane vertical velocities (bottom row) for three snapshots of our single stream simulation (Run 1). Spiral arms generated by the initial shock front of the streamer hitting the disk are seen in the first column and are labelled S1 and S2. Spiral arms intermediate to S1 and S2 are generated due to radially converging flows and are labelled S3. Spiral arms S1, S2, and S3, are highlighted with green dot-dashed, blue dashed, and solid purple lines, respectively. }
\label{fig:velocity}
\end{figure*}

\subsection{Infall Induced Spiral Arms and Velocity Residuals}

Figure~\ref{fig:velocity} plots the surface density, mid-plane radial velocities, and mid-plane vertical velocities (from top to bottom, respectively) for three snapshots of our simulation. The radial velocities are defined such that a positive radial velocity indicates material moving away from the center of the coordinate system, the star. Superimposed on the plot are points along the spiral arms obtained using the code {\sc nautilus}\footnote{\url{https://github.com/TomHilder/nautilus}}, with additional coloured solid, dashed, and dot-dashed lines also superimposed for clarity in later discussion. 

In the first snapshot (left column) two distinct spiral structures may be identified. These originate from the shock front of the infalling streamer, which we label S1 and S2 and highlight with green dot-dashed and blue dashed lines, respectively. The region in-between S1 and S3 is of comparatively lower density as a result of the infalling material colliding with the disk. We first focus on the radial velocities, which are mostly negative (moving towards the center) as the streamer is not quite at its pericenter passage before colliding with the disk. In the second snapshot (middle column), we see the appearance of additional spiral structures at the region between spirals S1 and S2, which we label S3 and highlight with a purple solid line. Although there are multiple spirals in the region, we label them all ``S3" since they appear to arise from the same origin. Looking at the radial velocities we see that the spirals S3 are at the interaction between colliding radially flows. The inner region of S3 contains radial flows with a positive sign (moving outwards), while the material outside of S3 has a negative sign (moving inward), creating an additional shock front that generates spiral structure. This radially converging flow persists only for a few orbits, and we see in the final column of Figure~\ref{fig:velocity} that the magnitude of the radial velocity flows has decreased by nearly a factor of two. 

With the radial velocity profiles shown in Figure~\ref{fig:velocity} and our previous qualitative description, it is straightforward to understand how infall can produce spiral arms with radially converging velocity similar to the gravitational instability. Infall introduces density perturbations away from hydrostatic equilibrium in the radial direction. In particular, the region in-between diverging shock fronts is of comparatively lower density than surrounding material. The disk responds with radially converging flows that attempt to re-fill this depletion of material and return to hydrostatic equilibrium. Therefore, radially converging flows around spiral arms are expected in disks with ongoing infall. 

We now focus on the vertical velocities measured at the midplane. Given that the streamer in the simulation is inclined with respect to the disk, substantial vertical velocity perturbations naturally arise. Of particular interest is the result that the vertical velocity perturbations are coherent across large azimuthal ranges. That is, large swathes of material across large azimuthal ranges spatially co-located with the spiral arms are moving in the same direction.

\subsection{Signatures in Molecular Line Observations}
\subsubsection{Channel Maps}
\begin{figure*}
\centering
\includegraphics[width=\textwidth]{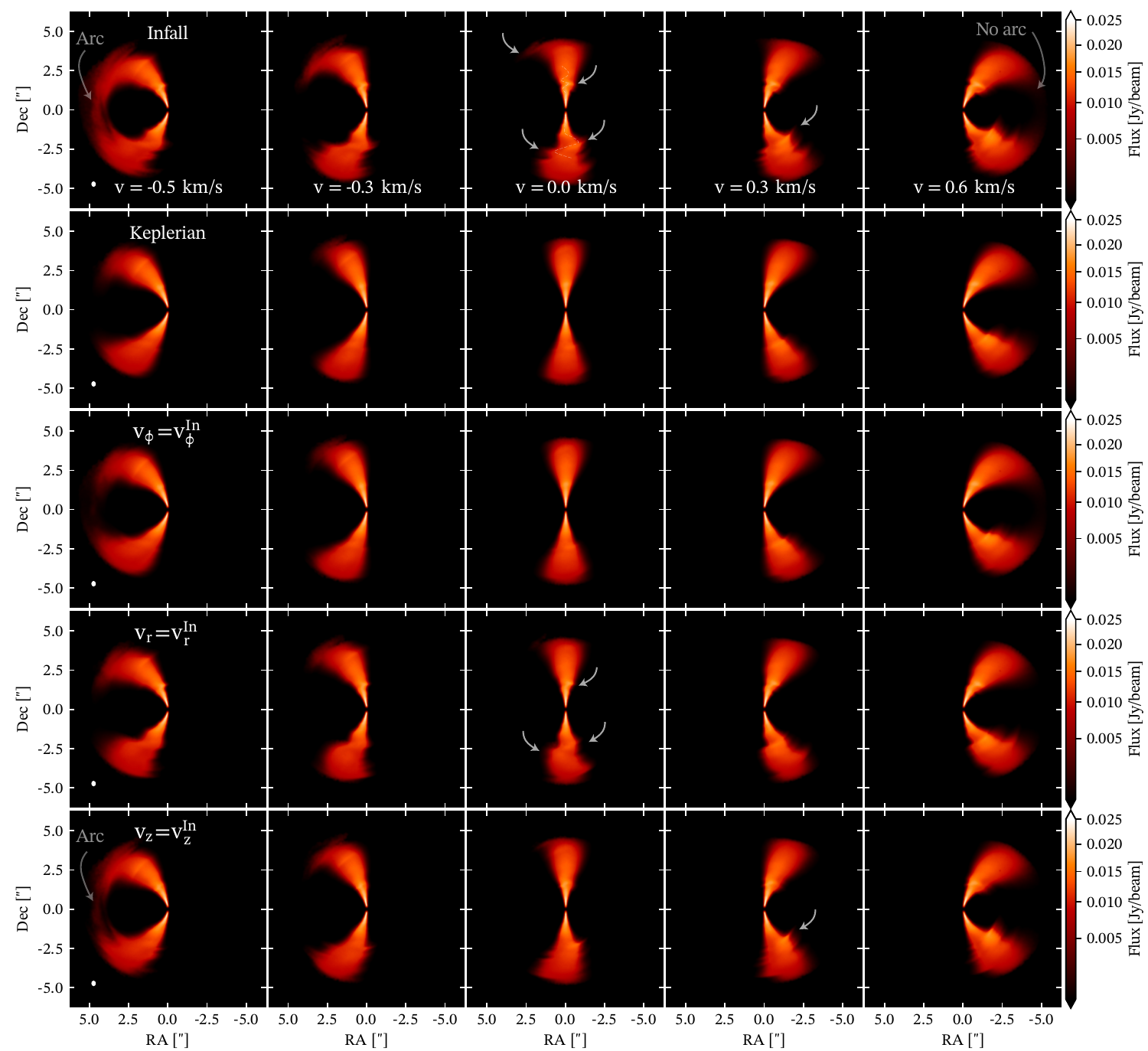}
\caption{The Infall Wiggle: $^{13}$CO channel maps from Run 1 made with different assumptions on the velocity field, produced at $t \sim 7250$ years (snapshot between the middle and right columns of Figure~\ref{fig:velocity}). The top row directly uses the radial ($v_r$), azimuthal ($v_\phi$), and vertical velocities ($v_z$) from the simulation. The second row assumes $v_\phi$ is at the Keplerian value and $v_r = v_z = 0$. The third row uses $v_\phi$ from the simulation, but $v_r = v_z = 0$. The fourth row uses $v_r$ from the simulation, with $v_z = 0$ and $v_\phi$ set to Keplerian. The final row assumes $v_z$ from the simulation, $v_r=0$, and $v_\phi$ is Keplerian.}
\label{fig:channels}
\end{figure*}

\begin{figure*}
\centering
\includegraphics[width=\textwidth]{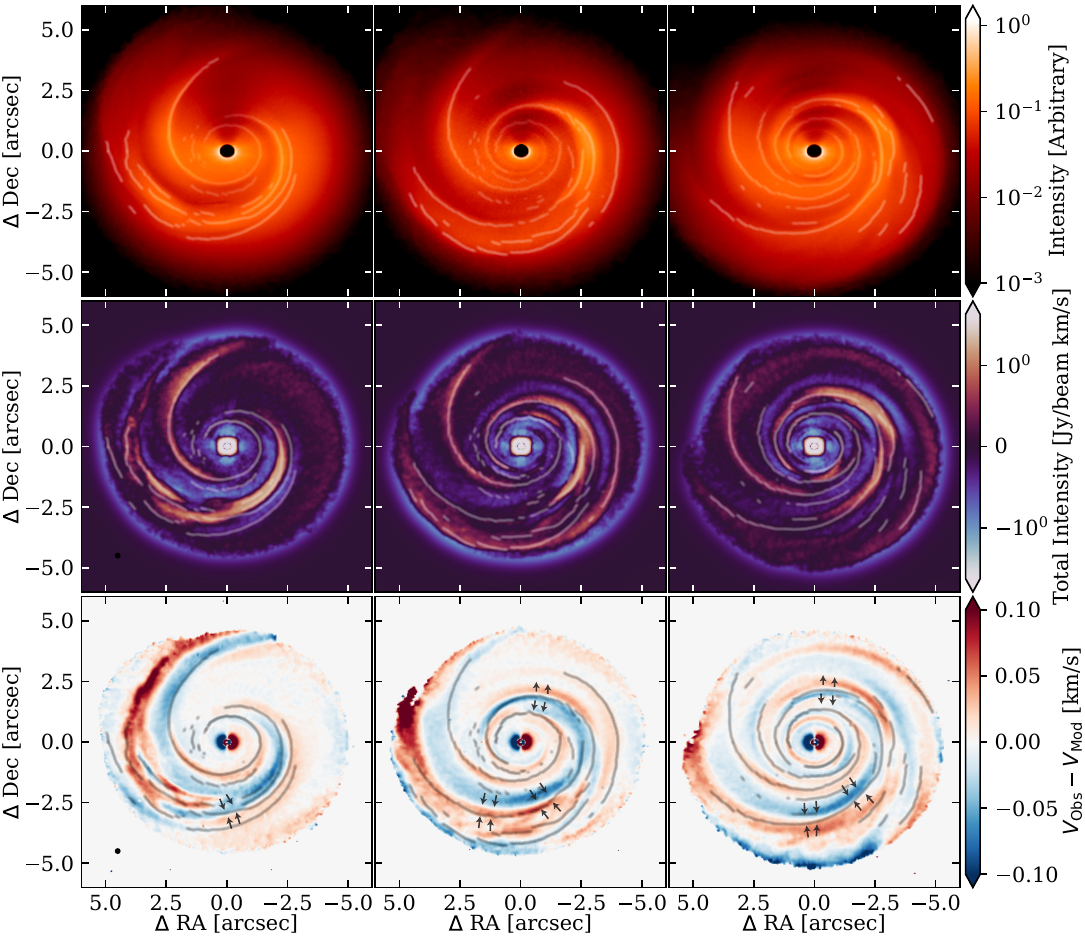}
\caption{Polarized scattered light intensity (top row), moment 0 residuals (middle row) and moment 1 residuals (bottom row) of three simulation snapshots at times $t = [5600, 7250, 8950]$ years. Residuals were obtained as in \citet{speedie2024}. Velocity residuals along the disk major axis show converging flows towards the location of spiral arms mapped in scattered light.}
\label{fig:residuals}
\end{figure*}

We assess which velocity component drives perturbations in the iso-velocity curves of our mock $^{13}$ CO (2-1) observations by building five sets of synthetic channel maps that differ only in the velocities assigned to the SPH particles:

\begin{enumerate}
    \item Full simulation: all three coordinate velocities taken directly from the hydrodynamic model.  
    \item Pure Keplerian rotation: $v_\phi = v_\phi^{\rm Kep}$, with $v_r = v_z = 0$.  
    \item Infall azimuthal only: $v_\phi = v_\phi^{\rm In}$, with $v_r = v_z = 0$.  
    \item Infall radial only: $v_r = v_r^{\rm In}$, while $v_\phi = v_\phi^{\rm Kep}$ and $v_z = 0$.  
    \item Infall vertical only: $v_z = v_z^{\rm In}$, with $v_\phi = v_\phi^{\rm Kep}$ and $v_r = 0$.
\end{enumerate}

The first row of Fig.~\ref{fig:channels} shows the full infall model, with the middle column showing the line-of-sight $v=0$ km/s channel, and the left and right of middle columns showing the blue and redshifted wings, respectively. ``Wiggles" in the iso-velocity curve are apparent across all the channel maps in the full simulation, with the middle column using arrows to signify particularly strong wiggles. The wiggles produced by infall are remarkably similar to those produced by the gravitational instability \citep{hall2020}. Also apparent is an ``arc''-like structure in the $v=-0.5$ km/s channel that connects the northern and southern wings of the iso-velocity curves. This arc structure is not apparent on the $v=0.5$ km/s channel, although smaller arcs are apparent in the $v=0.3$ km/s channel (marked with an arrow, fourth column). 

The second row of Figure~\ref{fig:channels} shows the channel maps assuming all the SPH particles are on circular Keplerian orbits. The iso-velocity curves show slight deviations from pure Keplerian ones, likely due to the changing height of the emitting layer as the under-lying density distribution is not smooth, and some regions are more optically thick than others. 
Using the azimuthal velocity values from the simulation (row 3) only slightly changes the iso-velocity curves compared with the pure Keplerian case. The large arc-like structure and wiggles in the $v=0$ km/s velocity channel are not visible and hence we conclude that the azimuthal velocity perturbations only play a minor role in their production. 

The fourth row shows the channel maps when we adopt the radial velocity values of the SPH simulations when running the radiative transfer model. The wiggles in the $v=0$ km/s channel map are recovered, as are additional wiggles across the different channels. Using only the vertical velocities from the SPH simulation, the arc in the $v=-0.6$ km/s channel becomes most visible, confirming that the azimuthally extended vertical velocity perturbations seen in the mid-plane of the simulation (Figure~\ref{fig:velocity}) are also present at the CO emitting layer.

\begin{figure*}
\centering
\includegraphics[width=\textwidth]{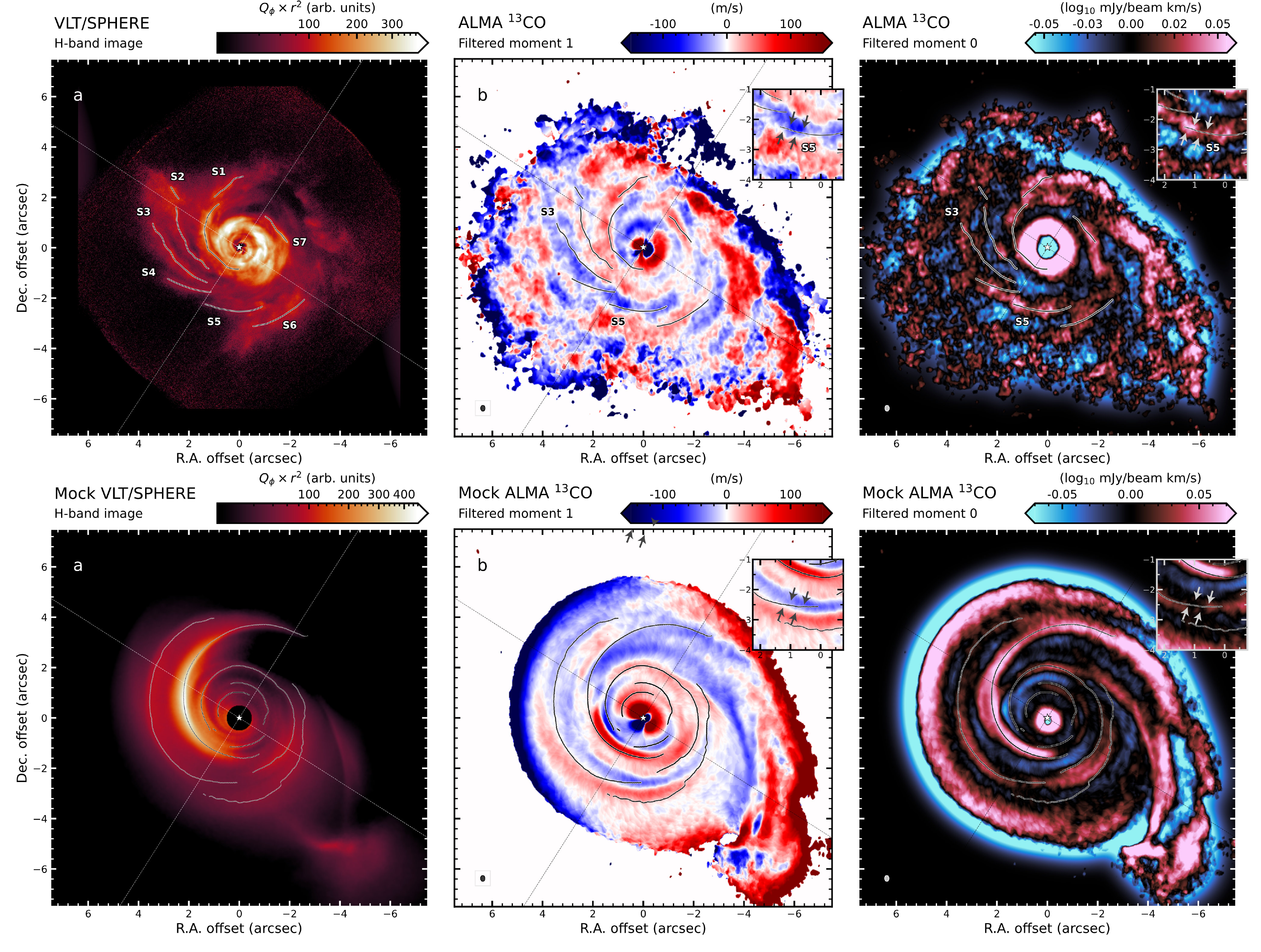}
\caption{The observations of AB~Aur with VLT/SPHERE and ALMA (top row) with the mock versions of our Run 2 shown at $t=6550$ yrs(bottom row). In both AB~Aur and our model, spiral arms seen in the scattered light correlate with patterns in the velocity residuals and spiral structures seen in the moment 0 residuals. The inset panels in the middle and right columns show the location of spiral structure close to the minor axis of the disk, where radial perturbations dominate over azimuthal perturbations. We see radially converging flows in our model that match those seen in AB~Aur.}
\label{fig:ab_residuals}
\end{figure*}

\subsubsection{Moment Maps}

In Figure~\ref{fig:residuals} we plot the $r^2$-scaled H-band polarized intensity scattered light image of our model in the top row, with residuals from moment 0 and moment 1 shown in the bottom panel, for three simulation snapshots at times $t = [5600, 7250, 8950]$ years. We set the color bar for our velocity residuals such that red indicates redshifted material and blue indicated blueshifted material. We produce our residual moment maps using the high-pass filtering with a radially expanding kernel technique to replicate the procedure from \cite{speedie2024}. The kernel is given by
\begin{equation}
    w(r) = w_0 (r/r_0)^\gamma,
\end{equation}
where $w_0$ is the kernel width at $r_0=1\arcsec$. We select $w_0=5$ pix (1.9 beams), $\gamma = 0.5$.

Spiral arms are visible in the polarized intensity and moment 0 residuals. Superimposed on each plot (white points in top two rows, black points in bottom) are the spiral structures identified with {\sc nautilus}. 
The location of the spiral arms in scattered light strongly correlates with the spiral arms in the moment 0 residuals, however they are not always directly on top of one another. This presents a challenge when interpreting the center of a particular spiral arm to then compare with the velocity residuals. However if we take at face value the spiral arm locations identified by {\sc nautilus}, we can see several locations where the velocity residuals show radially converging flows. When projected onto the sky, the observed velocity $v_\textrm{obs}$ is
\begin{align}
v_{\text{obs}} = &\; \underbrace{v_{\phi} \sin(i) \cos(\phi)}_{\text{rotational}} 
+ \underbrace{v_{r} \sin(i) \sin(\phi)}_{\text{radial}} \notag \\
& + \underbrace{v_{z} \cos(i)}_{\text{vertical}} 
+ v_{\text{systemic}}.
\end{align}
The contribution from the radial velocity component is largest when $\sin(\phi)=1$ which occurs along the minor axis of the disk. In observations we can therefore attribute velocity perturbations close to the minor axis as being predominantly radial. We see several locations, marked with arrows, where the velocity residuals show radially convergent flows towards the center of the spiral arm (arrows point towards the traced spiral arm). In conjunction with the radially convergent motion, we also see radially \emph{divergent} motion (arrows point away from the traced spiral arm), which is also seen in the spiral wake launched by a planet \citep{rafikov2002, bollati2021, calcino2023}. Both radially converging and radially diverging velocity flows are spatially co-located with spiral arms seen in scattered light and in CO moment 0 observations.

\subsection{Comparison with AB~Aur}
\begin{figure*}
\centering
\includegraphics[width=\textwidth]{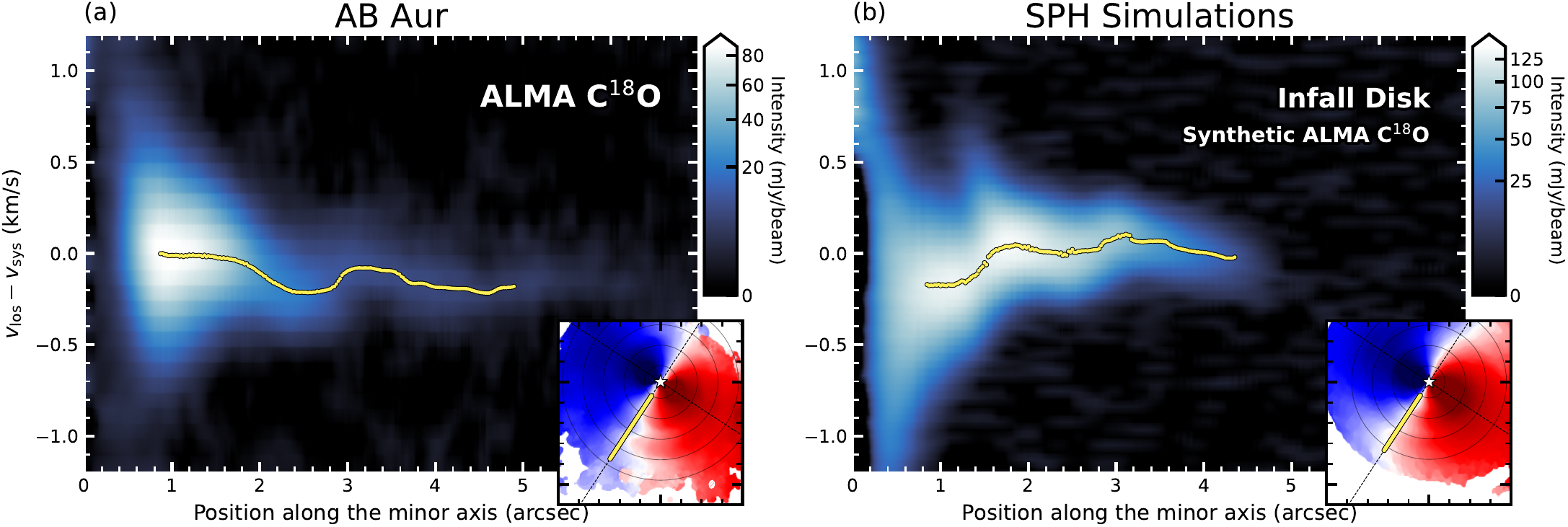}
\caption{The C$^{18}$O observations of AB~Aur in position-velocity space along the southern minor axis of the disk (left panel) compared with our infall simulation (right panel). Line centers are plotted with yellow points. Each panel contains an in inlay of the moment 1 which shows the location where the position-velocity data is extracted. }
\label{fig:ab_co_line}
\end{figure*}

Figure~\ref{fig:ab_residuals} shows the $r^2$-scaled SPHERE H-band polarized intensity observation of AB~Aur \citep{boccaletti2020}, which may be compared with the filtered moment 1 and moment 0 ALMA $^{13}$CO observations from \cite{speedie2024} (top row). The bottom row shows the simulated observations of Run 2 at $t\sim 11000$ years. The moment 0 and moment 1 residuals were processed using the same radially expanding kernel with $w_0 = 25$ ($\sim2.5$ beams) and $\gamma=0.25$. In the moment 1 residuals of AB~Aur, an inset axes shows a zoom-in of the radially converging velocity flows around the scattered light spiral labeled S5. It is also spatially co-located with excess emission in the moment 0 residuals (right panel). However, other spiral arms seen in scattered light are not coincident with radially converging flows along the disc minor axis. This is most evident in spiral S3, which shows diverging radial flows close to the minor axis.

For our simulated observations, we once again used {\sc nautilus} to trace the location of spiral arm structure in the polarized intensity image and super-impose these spirals on to the moment 1 and moment 0 residuals. In the moment 1 residual we can see similar radially converging flows towards the spiral arm seen in scattered light, also spatially co-located with excess emission in the moment 0 residuals.  Infall does not lead to a coherent set of spiral arm structures that always produce the same radial flow patterns. Indeed most spiral arm structures in scattered light are not coincident with radially converging flow patterns in the gas kinematics. This is in qualitative agreement with the scattered light and gas kinematics of AB~Aur.

Figure~\ref{fig:ab_co_line} shows the position-velocity diagram of a thin wedge along the southern minor axis of the disc in the observations of AB~Aur and our infall simulation. The left panel shows the C$^{18}$O line observations with the reported GI-induced ``PV wiggle'', predicted by analytic theory and simulations of GI-unstable discs \citep{Longarini2021, terry2022}. The right panel with infall onto a gravitational stable, low mass disk, also shows a very similar signature to the GI-induced wiggle, demonstrating such a signature is not unique to GI-unstable discs.

\section{Discussion}\label{sec:disc}

\subsection{Infall induced kinematic structures}

Our numerical experiments show that infall can produce several peculiar kinematic signatures in the channel maps and velocity residuals. Both radially converging and diverging flows are seen, which in the past have been used as hallmarks of either GI \citep[e.g.][]{hall2020} or planetary wakes \citep[e.g.][]{bollati2021, calcino2023}, respectively. In both GI and spiral arms launched by a planet, we expect the radial velocity signature to be spatially co-located with a peak in disk surface density, as spiral arms are surface density enhancements compared with the surrounding disk. Our simulations show that infall alone can also produce the same signatures.

The appearance of wiggles and arcs in the channel maps of our infall simulations is also reminiscent of the large kinematic perturbations seen in the disks of the exoALMA sample \citep{exo1}. \cite{exo10} identified large-scale non-Keplerian arcs, along with filaments connecting emitting layers. In MWC~758 they suggest that a large-scale non-Keplerian arc that appears outside of the disk may be due to the remnants of a stellar flyby \citep{cuello2020} or interaction with a cloud \citep[see also][]{winter2025}. Other disks in the sample, such as HD~135344B and J1604, show arcs within the disk, which may be due to buoyancy spiral generated by a planet \citep{bae2021}. Although our results show that such structures can be generated by streamer-like infall, thus far late-stage infall is not observed around these systems. Further investigation exploring longer timescales and more realistically seeded infalling material are required. 

\subsection{Infall and GI are not mutually exclusive}

Infall-induced kinematic perturbations are very similar to those produced by the GI. However, this result does not preclude the existence of GI in the outer regions of disks undergoing infall. Indeed infall is the likely driver of GI \citep{Longarini2025} and can amplify the amplitude of GI-generated spiral arms \citep{harsono2011}. The disk around AB~Aur may indeed be gravitationally unstable, but due to the observed and ongoing infall \citep{speedie2025}, kinematic evidence alone is not sufficient evidence to prove GI is the cause of the kinematic perturbations. Since infall produces significant deviations away from hydrostatic equilibrium, methods that rely on an underlying quasi-steady-state solution, such as using the disk rotation curve to measure disk mass \citep{Veronesi2021, lodato2023, Veronesi2024}, may also be susceptible to biased results due to infall. This should be tested.

A combination of observations across multiple wavelengths and tracing different disk components may be the key to deciding whether the disk around AB~Aur is gravitationally unstable. For example, near-infrared observations across decade long baselines could be used to constrain spiral arm rotation rates which can be compared with predictions of GI and infall models \citep[e.g.][]{ren2018, ren2020, xie2021, xie2023}. Observations at longer wavelengths can be used to search for evidence of dust trapping and growth within spiral arms, another observational signature suggestive of GI \citep{dipierro2014}.

\section{A Unified Model of AB~Aur} \label{sec:ab_aur_comp}

\begin{figure*}
\centering
\includegraphics[width=\textwidth]{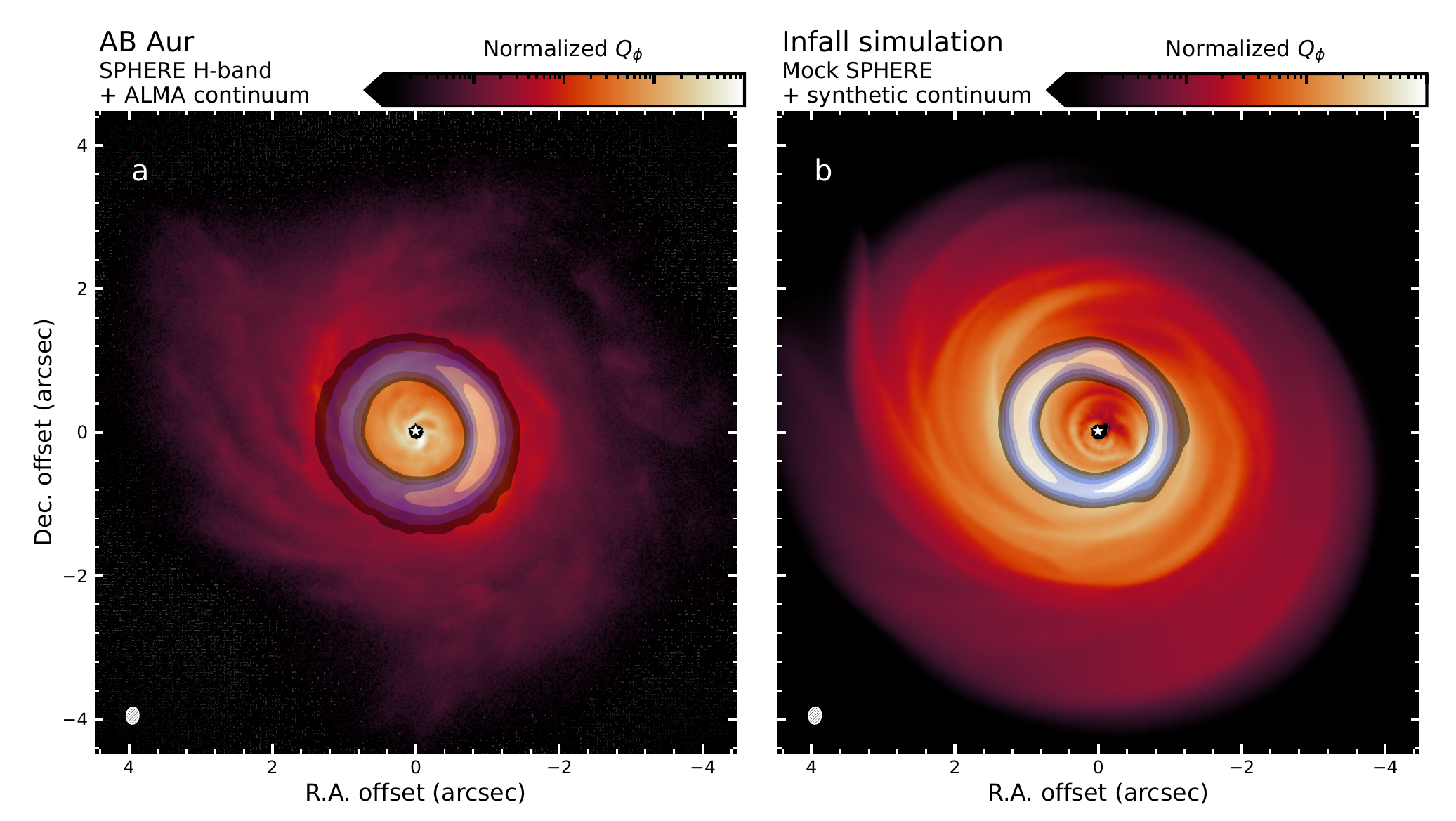}
\caption{A unified model for AB Aur. \emph{Left panel:} SPHERE H-band polarized intensity image of AB~Aur from \cite{boccaletti2020} and with ALMA dust continuum emission (Briggs robust 0.5 weighted) from \citet{speedie2024, speedie2025} super-imposed. \emph{Right panel:} The simulated H-band polarized intensity image of our dusty circumbinary disc with infall, with simulated ALMA band 6 mm dust continuum emission super-imposed.}
\label{fig:binary}
\end{figure*}

\cite{pietu2005}, \citet{tang2012-abaur-envelope,tang2017} and \citet{poblete2020} all argued for a low mass binary companion in the inner regions of AB Aur ($r \lesssim 0.5\arcsec$; $\sim 40$--$45$au), the former two studies from observations; the latter from a close match of simulations to observations in this region. Additional evidence from astrometric measurements by \citet{blakely2026}, and CO rovibrational spectroscopy by \citet{kozdon2026}, also support a massive companion at a similar separation. Binarity in disks with infall is expected on theoretical grounds: stars form in unstable multiple systems, so binaries are preferentially those that remain behind in the potential well where ongoing infall occurs \citep[e.g.][]{bate2003}. On statistical grounds alone, as a $\sim 2.4\,{\rm M}_\odot$ star \citep{tang2017}, AB Aur is 75\% likely to be multiple \citep{offner2023}. 

An inner massive companion in AB~Aur would be a natural explanation for the observed $^{13}$CO cavity, mm-dust continuum ring \citep{fuente2017}, and spiral structure within it \citep{tang2017}, as demonstrated by \citet{poblete2020}. We therefore evolved one additional model that includes an inner binary plus dust physics. The binary was modeled with sink particles of masses $M_1=2.3\,\textrm{M}_\odot$ and $M_2=0.2\,\textrm{M}_\odot$, with $a=54$ au, $e=0.5$, and $i=90^\circ$. The circumbinary disc was initialized with $5\times10^6$ particles between 120 and 600 au, with $R_c=420$ au, $\Sigma \propto R^{-1}$, a total mass of $M_\textrm{disk}=0.05\ $M$_\odot$, and $c_s\propto R^{-0.25}$, giving $H/R_\textrm{Ref}=0.06$ with $R_\textrm{ref}=120$ au. 

Dust was included using the multigrain algorithm \citep{laibe2014,price2015,multigrain2018} with 11 grain sizes between for $1\mu$m $\leq s \leq 1$mm with $dn/ds \propto s^{-3.5}$ and $\rho_\textrm{grain}=3\,{\rm g\,cm^{-3}}$. We included the dust-flux limiter from \cite{ballabio2018} to ensure mass conservation. Dust with a global dust-to-gas ratio of 100 was initially well-mixed with the gas, except for the largest four bins which were confined to the radial range between 120 and 180 au, consistent with the location of the dust ring seen in AB~Aur \citep{pietu2005,tang2012-abaur-envelope}. Since we only evolve the simulation for a few dozen dynamical timescales at the location of the dust ring and do not include dust grain growth, we assume that large dust grains have already grown and radially drifted towards the cavity edge. 

The system was evolved for 60 binary orbits before we added two streamers with masses of $2.5\times10^{-3}\,\textrm{M}_\odot$ each (total mass $\sim 10$\% the disk mass). The streamers were initialized as elliptical structures on parabolic orbits with $a=240$ au, $b=24$ au, $r_{\rm init}=720$ au, and $r_{\rm peri}=180$ au. They were inclined by $30^\circ$ and $50^\circ$, with an additional $20^\circ$ azimuthal rotation for the second. The dust-to-gas ratio inside the streamer was also set to 0.01, but only dust grains in the smallest size bin were included.

The snapshot used for radiative transfer was taken 22 binary orbits after the infall was added. When performing radiative transfer, we adopted a dust fluffiness parameter in {\sc mcfost} of 0.2, adjusting grain sizes and densities such that $\rho_{\textrm{grain}} s$, with $s$ being the grain size, remained constant resulting in the Stokes number also remaining unchanged. The secondary star was also included in the radiative transfer calculation, assuming $T_\star= 3800 K$, $R_\star=0.8\ $R$_\odot$, giving a luminosity of $L_\star = 0.08\ $L$_\odot$.  

Figure~\ref{fig:binary} shows the comparison between SPHERE H-band polarized intensity observations \citep{boccaletti2020} and ALMA band 6 continuum observations \citep{speedie2024} compared with the simulated observations of our dusty circumbinary disc with infall simulation. A 95 mas-in-radius coronograph is added as a simple mask in our mock SPHERE observations. The appearance of an asymmetric mm-dust horseshoe in our simulation is in line with previous studies that show such features can form in circumbinary discs \citep{ragusa2017, calcino2019, poblete2019, rabago2023}. While the dust horseshoe is not perfectly reproduced, simulations by \cite{rabago2023} show that vortices can form on the edge of the circumbinary disc cavity, also providing another explanation for the dust horseshoe in AB~Aur \citep{fuente2017}. 

The spiral arm structure in our model matches those seen in AB~Aur remarkably well. As noted by \citet{tang2012-abaur-envelope} the `multiple spirals [in the outer disk] are difficult to explain with a single perturbing body'. Furthermore, a companion inside the cavity would struggle to reproduce such spiral structure \citep[e.g. see][]{poblete2020, calcino2023}. Figure~\ref{fig:binary} demonstrates that most of the spiral arms in AB~Aur, particularly in the outer disc, can be naturally explained by infall. 

The inner cavity region of AB~Aur shows a higher polarized intensity than in our model. This is rather puzzling at first glance, since  the $^{13}$CO moment 0 strongly suggests a significant depletion of gas in this region. In some circumbinary discs, such as HD~142527, a cavity depleted in $^{13}$CO emission can also coincide with a cavity in polarized intensity \citep{avenhaus2014}. However not all circumbinary discs show this. For example, GG~Tau, whose central cavity contains a triple system \citep{difolco2014}, also shows significant scattering inside a gas depleted cavity \citep{Keppler2020}.  MWC~758 \citep{benisty2015, boehler2018} and CQ~Tau \citep{Uyama2020,wolfer2021} are two other discs that show spiral arm structures and significant scattering inside of a gas depleted cavity. 

Therefore, the lack of a cavity in scattered light in AB~Aur does not necessarily preclude the existence of a companion of stellar mass in this region. Perturbations directly from infall can act to refill the cavity with gas from the outer disc. In addition, \cite{speedie2025} also trace the tail-end of a streamer that appears to be colliding with the inner cavity region. Since streamers originate from the surrounding environment, it is reasonable to expect that their dust grain size distribution is ISM-like. An increase in small grains would act to increase the scattering opacity at optical and near-infrared wavelengths. Including this streamer in our simulation is difficult since it results in a high density of particles around each sink, drastically reducing the timestep. However it is a plausible explanation for the lack of a cavity in scattered light in AB~Aur. We also note that the stellar companion in our simulation would be hidden behind the 95 mas-in-radius coronograph in the observations at the snapshot shown.

\section{Summary}

Infall onto protoplanetary disks produce a variety of kinematic perturbations that can be mistaken for kinematic signatures of other phenomena. We have shown that infall can produce the following kinematic signatures in the moment 1 residual maps and channel maps:
\begin{enumerate}
    \item Radially flows that converge towards the location of spiral arms, similar to those produced by gravitational instability
    \item Radial flows that diverge from the location of spiral arms, similar to planetary wakes
    \item Wiggles in the iso-velocity curves of molecular line channel maps
    \item Large-scale non-Keplerian arcs of material
\end{enumerate}
The radially converging flows of infall induced spiral arms suggest that the kinematic features around the disk of AB~Aur \citep{speedie2024} may be explained by the observed infall \citep{speedie2025}.
Disentangling signatures of infall from those of other mechanisms, such as GI, is not trivial, and will likely require a combination of observations across multiple wavelengths and different tracers of disk properties. A combination of an inner central binary companion and infall onto the outer disk can explain the central gas and dust depleted cavity, mm dust continuum ring, abundant spiral structure, and perturbed kinematics in the disc around AB~Aur.

\section*{Data Availability}
The scripts and data required to reproduce all figures in this paper are available at \url{10.5281/zenodo.20266246}.

\section*{Acknowledgments}
We thank Jess Speedie for useful discussion, and the anonymous referee for their helpful suggestions that improved the quality of our manuscript. 
This work is supported by the National Natural Science Foundation of China under grant No. 12233004, 12250610189, and W2533008.
DJP is grateful for Australian Research Council funding via DP180104235 and DP220103767.
We used {\sc plonk} \citep{plonk2019} which utilizes functions written in {\sc splash} \citep{price2007}.
This work was performed on the OzSTAR national facility at Swinburne University of Technology. The OzSTAR program receives funding in part from the Astronomy National Collaborative Research Infrastructure Strategy (NCRIS) allocation provided by the Australian Government, and from the Victorian Higher Education State Investment Fund (VHESIF) provided by the Victorian Government. DJP thanks everyone in Grenoble for hospitality and support during his sabbatical visit in 2025.

\bibliography{paper}{}

\begin{thebibliography}{}
\expandafter\ifx\csname natexlab\endcsname\relax\def\natexlab#1{#1}\fi
\providecommand{\url}[1]{\href{#1}{#1}}
\providecommand{\dodoi}[1]{doi:~\href{http://doi.org/#1}{\nolinkurl{#1}}}
\providecommand{\doeprint}[1]{\href{http://ascl.net/#1}{\nolinkurl{http://ascl.net/#1}}}
\providecommand{\doarXiv}[1]{\href{https://arxiv.org/abs/#1}{\nolinkurl{https://arxiv.org/abs/#1}}}

\bibitem[{{Avenhaus} {et~al.}(2014){Avenhaus}, {Quanz}, {Schmid}, {Meyer}, {Garufi}, {Wolf}, \& {Dominik}}]{avenhaus2014}
{Avenhaus}, H., {Quanz}, S.~P., {Schmid}, H.~M., {et~al.} 2014, \apj, 781, 87, \dodoi{10.1088/0004-637X/781/2/87}

\bibitem[{{Bae} {et~al.}(2015){Bae}, {Hartmann}, \& {Zhu}}]{bae2015}
{Bae}, J., {Hartmann}, L., \& {Zhu}, Z. 2015, \apj, 805, 15, \dodoi{10.1088/0004-637X/805/1/15}

\bibitem[{{Bae} {et~al.}(2021){Bae}, {Teague}, \& {Zhu}}]{bae2021}
{Bae}, J., {Teague}, R., \& {Zhu}, Z. 2021, \apj, 912, 56, \dodoi{10.3847/1538-4357/abe45e}

\bibitem[{{Ballabio} {et~al.}(2018){Ballabio}, {Dipierro}, {Veronesi}, {Lodato}, {Hutchison}, {Laibe}, \& {Price}}]{ballabio2018}
{Ballabio}, G., {Dipierro}, G., {Veronesi}, B., {et~al.} 2018, \mnras, 477, 2766, \dodoi{10.1093/mnras/sty642}

\bibitem[{{Bate} {et~al.}(2003){Bate}, {Bonnell}, \& {Bromm}}]{bate2003}
{Bate}, M.~R., {Bonnell}, I.~A., \& {Bromm}, V. 2003, \mnras, 339, 577, \dodoi{10.1046/j.1365-8711.2003.06210.x}

\bibitem[{{Bate} {et~al.}(1995){Bate}, {Bonnell}, \& {Price}}]{bate1995}
{Bate}, M.~R., {Bonnell}, I.~A., \& {Price}, N.~M. 1995, \mnras, 277, 362, \dodoi{10.1093/mnras/277.2.362}

\bibitem[{{Benisty} {et~al.}(2015){Benisty}, {Juhasz}, {Boccaletti}, {Avenhaus}, {Milli}, {Thalmann}, {Dominik}, {Pinilla}, {Buenzli}, {Pohl}, {Beuzit}, {Birnstiel}, {de Boer}, {Bonnefoy}, {Chauvin}, {Christiaens}, {Garufi}, {Grady}, {Henning}, {Huelamo}, {Isella}, {Langlois}, {M{\'e}nard}, {Mouillet}, {Olofsson}, {Pantin}, {Pinte}, \& {Pueyo}}]{benisty2015}
{Benisty}, M., {Juhasz}, A., {Boccaletti}, A., {et~al.} 2015, \aap, 578, L6, \dodoi{10.1051/0004-6361/201526011}

\bibitem[{{Benisty} {et~al.}(2023){Benisty}, {Dominik}, {Follette}, {Garufi}, {Ginski}, {Hashimoto}, {Keppler}, {Kley}, \& {Monnier}}]{benisty2023}
{Benisty}, M., {Dominik}, C., {Follette}, K., {et~al.} 2023, in Astronomical Society of the Pacific Conference Series, Vol. 534, Protostars and Planets VII, ed. S.~{Inutsuka}, Y.~{Aikawa}, T.~{Muto}, K.~{Tomida}, \& M.~{Tamura}, 605, \dodoi{10.48550/arXiv.2203.09991}

\bibitem[{{Blakely} {et~al.}(2026){Blakely}, {Thompson}, {Johnstone}, {Speedie}, {Xuan}, {Blouin}, {Zhang}, {Ruffio}, {Nielsen}, {Bowler}, {Franson}, {Roberson}, {Cloutier}, {Fogal}, {Hessel}, {Marois}, \& {Rochon}}]{blakely2026}
{Blakely}, D., {Thompson}, W., {Johnstone}, D., {et~al.} 2026, arXiv e-prints, arXiv:2602.07731, \dodoi{10.48550/arXiv.2602.07731}

\bibitem[{{Boccaletti} {et~al.}(2020){Boccaletti}, {Di Folco}, {Pantin}, {Dutrey}, {Guilloteau}, {Tang}, {Pi{\'e}tu}, {Habart}, {Milli}, {Beck}, \& {Maire}}]{boccaletti2020}
{Boccaletti}, A., {Di Folco}, E., {Pantin}, E., {et~al.} 2020, \aap, 637, L5, \dodoi{10.1051/0004-6361/202038008}

\bibitem[{{Boehler} {et~al.}(2018){Boehler}, {Ricci}, {Weaver}, {Isella}, {Benisty}, {Carpenter}, {Grady}, {Shen}, {Tang}, \& {Perez}}]{boehler2018}
{Boehler}, Y., {Ricci}, L., {Weaver}, E., {et~al.} 2018, \apj, 853, 162, \dodoi{10.3847/1538-4357/aaa19c}

\bibitem[{{Bollati} {et~al.}(2021){Bollati}, {Lodato}, {Price}, \& {Pinte}}]{bollati2021}
{Bollati}, F., {Lodato}, G., {Price}, D.~J., \& {Pinte}, C. 2021, \mnras, 504, 5444, \dodoi{10.1093/mnras/stab1145}

\bibitem[{{Brittain} {et~al.}(2007){Brittain}, {Simon}, {Najita}, \& {Rettig}}]{brittain2007}
{Brittain}, S.~D., {Simon}, T., {Najita}, J.~R., \& {Rettig}, T.~W. 2007, \apj, 659, 685, \dodoi{10.1086/511255}

\bibitem[{{Cadman} {et~al.}(2021){Cadman}, {Rice}, \& {Hall}}]{cadman2021}
{Cadman}, J., {Rice}, K., \& {Hall}, C. 2021, \mnras, 504, 2877, \dodoi{10.1093/mnras/stab905}

\bibitem[{{Calcino} {et~al.}(2020){Calcino}, {Christiaens}, {Price}, {Pinte}, {Davis}, {van der Marel}, \& {Cuello}}]{calcino2020}
{Calcino}, J., {Christiaens}, V., {Price}, D.~J., {et~al.} 2020, \mnras, 498, 639, \dodoi{10.1093/mnras/staa2468}

\bibitem[{{Calcino} {et~al.}(2022){Calcino}, {Hilder}, {Price}, {Pinte}, {Bollati}, {Lodato}, \& {Norfolk}}]{calcino2022}
{Calcino}, J., {Hilder}, T., {Price}, D.~J., {et~al.} 2022, \apjl, 929, L25, \dodoi{10.3847/2041-8213/ac64a7}

\bibitem[{{Calcino} {et~al.}(2025){Calcino}, {Price}, {Hilder}, {Christiaens}, {Speedie}, \& {Ormel}}]{calcino2025}
{Calcino}, J., {Price}, D.~J., {Hilder}, T., {et~al.} 2025, \mnras, 537, 2695, \dodoi{10.1093/mnras/staf135}

\bibitem[{{Calcino} {et~al.}(2023){Calcino}, {Price}, {Pinte}, {Garg}, {Norfolk}, {Christiaens}, {Li}, \& {Teague}}]{calcino2023}
{Calcino}, J., {Price}, D.~J., {Pinte}, C., {et~al.} 2023, \mnras, 523, 5763, \dodoi{10.1093/mnras/stad1798}

\bibitem[{{Calcino} {et~al.}(2019){Calcino}, {Price}, {Pinte}, {van der Marel}, {Ragusa}, {Dipierro}, {Cuello}, \& {Christiaens}}]{calcino2019}
---. 2019, \mnras, 490, 2579, \dodoi{10.1093/mnras/stz2770}

\bibitem[{{Calcino} {et~al.}(2024){Calcino}, {Norfolk}, {Price}, {Hilder}, {Speedie}, {Pinte}, {Garg}, {Teague}, {Hall}, \& {Stadler}}]{calcino2024}
{Calcino}, J., {Norfolk}, B.~J., {Price}, D.~J., {et~al.} 2024, \mnras, 534, 2904, \dodoi{10.1093/mnras/stae2233}

\bibitem[{{Cuello} {et~al.}(2020){Cuello}, {Louvet}, {Mentiplay}, {Pinte}, {Price}, {Winter}, {Nealon}, {M{\'e}nard}, {Lodato}, {Dipierro}, {Christiaens}, {Montesinos}, {Cuadra}, {Laibe}, {Cieza}, {Dong}, \& {Alexander}}]{cuello2020}
{Cuello}, N., {Louvet}, F., {Mentiplay}, D., {et~al.} 2020, \mnras, 491, 504, \dodoi{10.1093/mnras/stz2938}

\bibitem[{{Di Folco} {et~al.}(2014){Di Folco}, {Dutrey}, {Le Bouquin}, {Lacour}, {Berger}, {K{\"o}hler}, {Guilloteau}, {Pi{\'e}tu}, {Bary}, {Beck}, {Beust}, \& {Pantin}}]{difolco2014}
{Di Folco}, E., {Dutrey}, A., {Le Bouquin}, J.~B., {et~al.} 2014, \aap, 565, L2, \dodoi{10.1051/0004-6361/201423675}

\bibitem[{{Dipierro} {et~al.}(2014){Dipierro}, {Lodato}, {Testi}, \& {de Gregorio Monsalvo}}]{dipierro2014}
{Dipierro}, G., {Lodato}, G., {Testi}, L., \& {de Gregorio Monsalvo}, I. 2014, \mnras, 444, 1919, \dodoi{10.1093/mnras/stu1584}

\bibitem[{{Dipierro} {et~al.}(2015){Dipierro}, {Price}, {Laibe}, {Hirsh}, {Cerioli}, \& {Lodato}}]{dipierro2015}
{Dipierro}, G., {Price}, D., {Laibe}, G., {et~al.} 2015, \mnras, 453, L73, \dodoi{10.1093/mnrasl/slv105}

\bibitem[{{Donehew} \& {Brittain}(2011)}]{donehew2011}
{Donehew}, B., \& {Brittain}, S. 2011, \aj, 141, 46, \dodoi{10.1088/0004-6256/141/2/46}

\bibitem[{{Dullemond} {et~al.}(2019){Dullemond}, {K{\"u}ffmeier}, {Goicovic}, {Fukagawa}, {Oehl}, \& {Kramer}}]{dullemond2019}
{Dullemond}, C.~P., {K{\"u}ffmeier}, M., {Goicovic}, F., {et~al.} 2019, \aap, 628, A20, \dodoi{10.1051/0004-6361/201832632}

\bibitem[{{Flores} {et~al.}(2023){Flores}, {Ohashi}, {Tobin}, {J{\o}rgensen}, {Takakuwa}, {Li}, {Lin}, {van't Hoff}, {Plunkett}, {Yamato}, {Sai (Insa Choi)}, {Koch}, {Yen}, {Aikawa}, {Aso}, {de Gregorio-Monsalvo}, {Kido}, {Kwon}, {Lee}, {Lee}, {Looney}, {Santamar{\'\i}a-Miranda}, {Sharma}, {Thieme}, {Williams}, {Han}, {Narayanan}, \& {Lai}}]{flores2023}
{Flores}, C., {Ohashi}, N., {Tobin}, J.~J., {et~al.} 2023, \apj, 958, 98, \dodoi{10.3847/1538-4357/acf7c1}

\bibitem[{{Fuente} {et~al.}(2017){Fuente}, {Baruteau}, {Neri}, {Carmona}, {Ag{\'u}ndez}, {Goicoechea}, {Bachiller}, {Cernicharo}, \& {Bern{\'e}}}]{fuente2017}
{Fuente}, A., {Baruteau}, C., {Neri}, R., {et~al.} 2017, \apjl, 846, L3, \dodoi{10.3847/2041-8213/aa8558}

\bibitem[{{Fukagawa} {et~al.}(2004){Fukagawa}, {Hayashi}, {Tamura}, {Itoh}, {Hayashi}, {Oasa}, {Takeuchi}, {Morino}, {Murakawa}, {Oya}, {Yamashita}, {Suto}, {Mayama}, {Naoi}, {Ishii}, {Pyo}, {Nishikawa}, {Takato}, {Usuda}, {Ando}, {Iye}, {Miyama}, \& {Kaifu}}]{fukagawa2004}
{Fukagawa}, M., {Hayashi}, M., {Tamura}, M., {et~al.} 2004, \apjl, 605, L53, \dodoi{10.1086/420699}

\bibitem[{{Garcia Lopez} {et~al.}(2006){Garcia Lopez}, {Natta}, {Testi}, \& {Habart}}]{Garcia-Lopez+2006}
{Garcia Lopez}, R., {Natta}, A., {Testi}, L., \& {Habart}, E. 2006, \aap, 459, 837, \dodoi{10.1051/0004-6361:20065575}

\bibitem[{{Garufi} {et~al.}(2022){Garufi}, {Podio}, {Codella}, {Segura-Cox}, {Vander Donckt}, {Mercimek}, {Bacciotti}, {Fedele}, {Kasper}, {Pineda}, {Humphreys}, \& {Testi}}]{garufi2022}
{Garufi}, A., {Podio}, L., {Codella}, C., {et~al.} 2022, \aap, 658, A104, \dodoi{10.1051/0004-6361/202141264}

\bibitem[{{Garufi} {et~al.}(2024){Garufi}, {Ginski}, {van Holstein}, {Benisty}, {Manara}, {P{\'e}rez}, {Pinilla}, {Ribas}, {Weber}, {Williams}, {Cieza}, {Dominik}, {Facchini}, {Huang}, {Zurlo}, {Bae}, {Hagelberg}, {Henning}, {Hogerheijde}, {Janson}, {M{\'e}nard}, {Messina}, {Meyer}, {Pinte}, {Quanz}, {Rigliaco}, {Roccatagliata}, {Schmid}, {Szul{\'a}gyi}, {van Boekel}, {Wahhaj}, {Antichi}, {Baruffolo}, \& {Moulin}}]{garufi2024}
{Garufi}, A., {Ginski}, C., {van Holstein}, R.~G., {et~al.} 2024, \aap, 685, A53, \dodoi{10.1051/0004-6361/202347586}

\bibitem[{{Goodman} \& {Rafikov}(2001)}]{goodman2001}
{Goodman}, J., \& {Rafikov}, R.~R. 2001, \apj, 552, 793, \dodoi{10.1086/320572}

\bibitem[{{Gupta} {et~al.}(2024){Gupta}, {Miotello}, {Williams}, {Birnstiel}, {Kuffmeier}, \& {Yen}}]{gupta2024}
{Gupta}, A., {Miotello}, A., {Williams}, J.~P., {et~al.} 2024, \aap, 683, A133, \dodoi{10.1051/0004-6361/202348007}

\bibitem[{{Hall} {et~al.}(2020){Hall}, {Dong}, {Teague}, {Terry}, {Pinte}, {Paneque-Carre{\~n}o}, {Veronesi}, {Alexander}, \& {Lodato}}]{hall2020}
{Hall}, C., {Dong}, R., {Teague}, R., {et~al.} 2020, \apj, 904, 148, \dodoi{10.3847/1538-4357/abac17}

\bibitem[{{Harsono} {et~al.}(2011){Harsono}, {Alexander}, \& {Levin}}]{harsono2011}
{Harsono}, D., {Alexander}, R.~D., \& {Levin}, Y. 2011, \mnras, 413, 423, \dodoi{10.1111/j.1365-2966.2010.18146.x}

\bibitem[{{Huang} {et~al.}(2020){Huang}, {Andrews}, {{\"O}berg}, {Ansdell}, {Benisty}, {Carpenter}, {Isella}, {P{\'e}rez}, {Ricci}, {Williams}, {Wilner}, \& {Zhu}}]{huang2020}
{Huang}, J., {Andrews}, S.~M., {{\"O}berg}, K.~I., {et~al.} 2020, \apj, 898, 140, \dodoi{10.3847/1538-4357/aba1e1}

\bibitem[{{Huang} {et~al.}(2021){Huang}, {Bergin}, {{\"O}berg}, {Andrews}, {Teague}, {Law}, {Kalas}, {Aikawa}, {Bae}, {Bergner}, {Booth}, {Bosman}, {Calahan}, {Cataldi}, {Cleeves}, {Czekala}, {Ilee}, {Le Gal}, {Guzm{\'a}n}, {Long}, {Loomis}, {M{\'e}nard}, {Nomura}, {Qi}, {Schwarz}, {Tsukagoshi}, {van't Hoff}, {Walsh}, {Wilner}, {Yamato}, \& {Zhang}}]{huang2021}
{Huang}, J., {Bergin}, E.~A., {{\"O}berg}, K.~I., {et~al.} 2021, \apjs, 257, 19, \dodoi{10.3847/1538-4365/ac143e}

\bibitem[{{H{\"u}hn} {et~al.}(2026){H{\"u}hn}, {Kimmig}, \& {Dullemond}}]{huhn2026}
{H{\"u}hn}, L.-A., {Kimmig}, C.~N., \& {Dullemond}, C.~P. 2026, \aap, 708, A93, \dodoi{10.1051/0004-6361/202558773}

\bibitem[{{Hutchison} {et~al.}(2018){Hutchison}, {Price}, \& {Laibe}}]{multigrain2018}
{Hutchison}, M., {Price}, D.~J., \& {Laibe}, G. 2018, \mnras, 476, 2186, \dodoi{10.1093/mnras/sty367}

\bibitem[{{Kama} {et~al.}(2020){Kama}, {Trapman}, {Fedele}, {Bruderer}, {Hogerheijde}, {Miotello}, {van Dishoeck}, {Clarke}, \& {Bergin}}]{kama2020}
{Kama}, M., {Trapman}, L., {Fedele}, D., {et~al.} 2020, \aap, 634, A88, \dodoi{10.1051/0004-6361/201937124}

\bibitem[{{Keppler} {et~al.}(2020){Keppler}, {Penzlin}, {Benisty}, {van Boekel}, {Henning}, {van Holstein}, {Kley}, {Garufi}, {Ginski}, {Brandner}, {Bertrang}, {Boccaletti}, {de Boer}, {Bonavita}, {Brown Sevilla}, {Chauvin}, {Dominik}, {Janson}, {Langlois}, {Lodato}, {Maire}, {M{\'e}nard}, {Pantin}, {Pinte}, {Stolker}, {Szul{\'a}gyi}, {Thebault}, {Villenave}, {Zurlo}, {Rabou}, {Feautrier}, {Feldt}, {Madec}, \& {Wildi}}]{Keppler2020}
{Keppler}, M., {Penzlin}, A., {Benisty}, M., {et~al.} 2020, \aap, 639, A62, \dodoi{10.1051/0004-6361/202038032}

\bibitem[{{Kozdon} {et~al.}(2026){Kozdon}, {Fung}, {Brittain}, {Jensen}, {Kern}, {Padgett}, \& {Hasegawa}}]{kozdon2026}
{Kozdon}, J., {Fung}, J., {Brittain}, S.~D., {et~al.} 2026, \aj, 171, 250, \dodoi{10.3847/1538-3881/ae48f6}

\bibitem[{{Kuffmeier} {et~al.}(2021){Kuffmeier}, {Dullemond}, {Reissl}, \& {Goicovic}}]{kuffmeier2021}
{Kuffmeier}, M., {Dullemond}, C.~P., {Reissl}, S., \& {Goicovic}, F.~G. 2021, \aap, 656, A161, \dodoi{10.1051/0004-6361/202039614}

\bibitem[{{Kuffmeier} {et~al.}(2018){Kuffmeier}, {Frimann}, {Jensen}, \& {Haugb{\o}lle}}]{kuffmeier2018b}
{Kuffmeier}, M., {Frimann}, S., {Jensen}, S.~S., \& {Haugb{\o}lle}, T. 2018, \mnras, 475, 2642, \dodoi{10.1093/mnras/sty024}

\bibitem[{{Kuznetsova} {et~al.}(2022){Kuznetsova}, {Bae}, {Hartmann}, \& {Mac Low}}]{Kuznetsova2022}
{Kuznetsova}, A., {Bae}, J., {Hartmann}, L., \& {Mac Low}, M.-M. 2022, \apj, 928, 92, \dodoi{10.3847/1538-4357/ac54a8}

\bibitem[{{Laibe} \& {Price}(2014)}]{laibe2014}
{Laibe}, G., \& {Price}, D.~J. 2014, \mnras, 440, 2136, \dodoi{10.1093/mnras/stu355}

\bibitem[{{Lodato} \& {Price}(2010)}]{lodato2010}
{Lodato}, G., \& {Price}, D.~J. 2010, \mnras, 405, 1212, \dodoi{10.1111/j.1365-2966.2010.16526.x}

\bibitem[{{Lodato} \& {Rice}(2005)}]{lodato2005}
{Lodato}, G., \& {Rice}, W.~K.~M. 2005, \mnras, 358, 1489, \dodoi{10.1111/j.1365-2966.2005.08875.x}

\bibitem[{{Lodato} {et~al.}(2023){Lodato}, {Rampinelli}, {Viscardi}, {Longarini}, {Izquierdo}, {Paneque-Carre{\~n}o}, {Testi}, {Facchini}, {Miotello}, {Veronesi}, \& {Hall}}]{lodato2023}
{Lodato}, G., {Rampinelli}, L., {Viscardi}, E., {et~al.} 2023, \mnras, 518, 4481, \dodoi{10.1093/mnras/stac3223}

\bibitem[{{Longarini} {et~al.}(2021){Longarini}, {Lodato}, {Toci}, {Veronesi}, {Hall}, {Dong}, \& {Patrick Terry}}]{Longarini2021}
{Longarini}, C., {Lodato}, G., {Toci}, C., {et~al.} 2021, \apjl, 920, L41, \dodoi{10.3847/2041-8213/ac2df6}

\bibitem[{{Longarini} {et~al.}(2025){Longarini}, {Price}, {Kratter}, {Lodato}, \& {Clarke}}]{Longarini2025}
{Longarini}, C., {Price}, D.~J., {Kratter}, K.~M., {Lodato}, G., \& {Clarke}, C.~J. 2025, \mnras, 541, 1145, \dodoi{10.1093/mnras/staf1018}

\bibitem[{{Mentiplay}(2019)}]{plonk2019}
{Mentiplay}, D. 2019, The Journal of Open Source Software, 4, 1884, \dodoi{10.21105/joss.01884}

\bibitem[{{Mesa} {et~al.}(2022){Mesa}, {Ginski}, {Gratton}, {Ertel}, {Wagner}, {Bonavita}, {Fedele}, {Meyer}, {Henning}, {Langlois}, {Garufi}, {Antoniucci}, {Claudi}, {Defr{\`e}re}, {Desidera}, {Janson}, {Pawellek}, {Rigliaco}, {Squicciarini}, {Zurlo}, {Boccaletti}, {Bonnefoy}, {Cantalloube}, {Chauvin}, {Feldt}, {Hagelberg}, {Hugot}, {Lagrange}, {Lazzoni}, {Maurel}, {Perrot}, {Petit}, {Rouan}, \& {Vigan}}]{Mesa2022}
{Mesa}, D., {Ginski}, C., {Gratton}, R., {et~al.} 2022, \aap, 658, A63, \dodoi{10.1051/0004-6361/202142219}

\bibitem[{{Milam} {et~al.}(2005){Milam}, {Savage}, {Brewster}, {Ziurys}, \& {Wyckoff}}]{milam2005}
{Milam}, S.~N., {Savage}, C., {Brewster}, M.~A., {Ziurys}, L.~M., \& {Wyckoff}, S. 2005, \apj, 634, 1126, \dodoi{10.1086/497123}

\bibitem[{{Monaghan}(1992)}]{monaghan1992}
{Monaghan}, J.~J. 1992, \araa, 30, 543, \dodoi{10.1146/annurev.aa.30.090192.002551}

\bibitem[{{Norfolk} {et~al.}(2022){Norfolk}, {Pinte}, {Calcino}, {Hammond}, {van der Marel}, {Price}, {Maddison}, {Christiaens}, {Gonzalez}, {Blakely}, {Rosotti}, \& {Ginski}}]{norfolk2022}
{Norfolk}, B.~J., {Pinte}, C., {Calcino}, J., {et~al.} 2022, \apjl, 936, L4, \dodoi{10.3847/2041-8213/ac85ed}

\bibitem[{{Offner} {et~al.}(2023){Offner}, {Moe}, {Kratter}, {Sadavoy}, {Jensen}, \& {Tobin}}]{offner2023}
{Offner}, S.~S.~R., {Moe}, M., {Kratter}, K.~M., {et~al.} 2023, in Astronomical Society of the Pacific Conference Series, Vol. 534, Protostars and Planets VII, ed. S.~{Inutsuka}, Y.~{Aikawa}, T.~{Muto}, K.~{Tomida}, \& M.~{Tamura}, 275, \dodoi{10.48550/arXiv.2203.10066}

\bibitem[{{Offner} {et~al.}(2012){Offner}, {Robitaille}, {Hansen}, {McKee}, \& {Klein}}]{offner2012}
{Offner}, S. S.~R., {Robitaille}, T.~P., {Hansen}, C.~E., {McKee}, C.~F., \& {Klein}, R.~I. 2012, \apj, 753, 98, \dodoi{10.1088/0004-637X/753/2/98}

\bibitem[{{Pelkonen} {et~al.}(2025){Pelkonen}, {Padoan}, {Juvela}, {Haugb{\o}lle}, \& {Nordlund}}]{pelkonen2025}
{Pelkonen}, V.-M., {Padoan}, P., {Juvela}, M., {Haugb{\o}lle}, T., \& {Nordlund}, {\r{A}}. 2025, \aap, 694, A327, \dodoi{10.1051/0004-6361/202450682}

\bibitem[{{P{\'e}rez} {et~al.}(2018){P{\'e}rez}, {Casassus}, \& {Ben{\'\i}tez-Llambay}}]{perez2018}
{P{\'e}rez}, S., {Casassus}, S., \& {Ben{\'\i}tez-Llambay}, P. 2018, \mnras, 480, L12, \dodoi{10.1093/mnrasl/sly109}

\bibitem[{{Pi{\'e}tu} {et~al.}(2005){Pi{\'e}tu}, {Guilloteau}, \& {Dutrey}}]{pietu2005}
{Pi{\'e}tu}, V., {Guilloteau}, S., \& {Dutrey}, A. 2005, \aap, 443, 945, \dodoi{10.1051/0004-6361:20042050}

\bibitem[{{Pinte} {et~al.}(2009){Pinte}, {Harries}, {Min}, {Watson}, {Dullemond}, {Woitke}, {M{\'e}nard}, \& {Dur{\'a}n-Rojas}}]{pinte2009}
{Pinte}, C., {Harries}, T.~J., {Min}, M., {et~al.} 2009, \aap, 498, 967, \dodoi{10.1051/0004-6361/200811555}

\bibitem[{{Pinte} {et~al.}(2006){Pinte}, {M{\'e}nard}, {Duch{\^e}ne}, \& {Bastien}}]{pinte2006}
{Pinte}, C., {M{\'e}nard}, F., {Duch{\^e}ne}, G., \& {Bastien}, P. 2006, \aap, 459, 797, \dodoi{10.1051/0004-6361:20053275}

\bibitem[{{Pinte} {et~al.}(2023){Pinte}, {Teague}, {Flaherty}, {Hall}, {Facchini}, \& {Casassus}}]{pinte2023}
{Pinte}, C., {Teague}, R., {Flaherty}, K., {et~al.} 2023, in Astronomical Society of the Pacific Conference Series, Vol. 534, Protostars and Planets VII, ed. S.~{Inutsuka}, Y.~{Aikawa}, T.~{Muto}, K.~{Tomida}, \& M.~{Tamura}, 645, \dodoi{10.48550/arXiv.2203.09528}

\bibitem[{{Pinte} {et~al.}(2018{\natexlab{a}}){Pinte}, {Price}, {M{\'e}nard}, {Duch{\^e}ne}, {Dent}, {Hill}, {de Gregorio-Monsalvo}, {Hales}, \& {Mentiplay}}]{pinte2018}
{Pinte}, C., {Price}, D.~J., {M{\'e}nard}, F., {et~al.} 2018{\natexlab{a}}, \apjl, 860, L13, \dodoi{10.3847/2041-8213/aac6dc}

\bibitem[{{Pinte} {et~al.}(2018{\natexlab{b}}){Pinte}, {M{\'e}nard}, {Duch{\^e}ne}, {Hill}, {Dent}, {Woitke}, {Maret}, {van der Plas}, {Hales}, {Kamp}, {Thi}, {de Gregorio-Monsalvo}, {Rab}, {Quanz}, {Avenhaus}, {Carmona}, \& {Casassus}}]{pinte2018b}
{Pinte}, C., {M{\'e}nard}, F., {Duch{\^e}ne}, G., {et~al.} 2018{\natexlab{b}}, \aap, 609, A47, \dodoi{10.1051/0004-6361/201731377}

\bibitem[{{Pinte} {et~al.}(2025){Pinte}, {Ilee}, {Huang}, {Benisty}, {Facchini}, {Fukagawa}, {Teague}, {Bae}, {Barraza-Alfaro}, {Cataldi}, {Cuello}, {Curone}, {Czekala}, {Fasano}, {Flock}, {Galloway-Sprietsma}, {Garg}, {Hall}, {Hammond}, {Hardiman}, {Hilder}, {Izquierdo}, {Kanagawa}, {Lesur}, {Lodato}, {Longarini}, {Loomis}, {Masset}, {Menard}, {Orihara}, {Price}, {Rosotti}, {Stadler}, {Yen}, {Wafflard-Fernandez}, {Wilner}, {Winter}, {W{\"o}lfer}, {Yoshida}, \& {Zawadzki}}]{exo10}
{Pinte}, C., {Ilee}, J.~D., {Huang}, J., {et~al.} 2025, \apjl, 984, L15, \dodoi{10.3847/2041-8213/adc433}

\bibitem[{{Poblete} {et~al.}(2020){Poblete}, {Calcino}, {Cuello}, {Mac{\'\i}as}, {Ribas}, {Price}, {Cuadra}, \& {Pinte}}]{poblete2020}
{Poblete}, P.~P., {Calcino}, J., {Cuello}, N., {et~al.} 2020, \mnras, \dodoi{10.1093/mnras/staa1655}

\bibitem[{{Poblete} {et~al.}(2019){Poblete}, {Cuello}, \& {Cuadra}}]{poblete2019}
{Poblete}, P.~P., {Cuello}, N., \& {Cuadra}, J. 2019, \mnras, 489, 2204, \dodoi{10.1093/mnras/stz2297}

\bibitem[{{Price}(2007)}]{price2007}
{Price}, D.~J. 2007, \pasa, 24, 159, \dodoi{10.1071/AS07022}

\bibitem[{{Price}(2012)}]{price2012}
---. 2012, Journal of Computational Physics, 231, 759, \dodoi{10.1016/j.jcp.2010.12.011}

\bibitem[{{Price} \& {Laibe}(2015)}]{price2015}
{Price}, D.~J., \& {Laibe}, G. 2015, \mnras, 451, 813, \dodoi{10.1093/mnras/stv996}

\bibitem[{{Price} {et~al.}(2018{\natexlab{a}}){Price}, {Cuello}, {Pinte}, {Mentiplay}, {Casassus}, {Christiaens}, {Kennedy}, {Cuadra}, {Sebastian Perez}, {Marino}, {Armitage}, {Zurlo}, {Juhasz}, {Ragusa}, {Laibe}, \& {Lodato}}]{price2018}
{Price}, D.~J., {Cuello}, N., {Pinte}, C., {et~al.} 2018{\natexlab{a}}, \mnras, 477, 1270, \dodoi{10.1093/mnras/sty647}

\bibitem[{{Price} {et~al.}(2018{\natexlab{b}}){Price}, {Wurster}, {Tricco}, {Nixon}, {Toupin}, {Pettitt}, {Chan}, {Mentiplay}, {Laibe}, {Glover}, {Dobbs}, {Nealon}, {Liptai}, {Worpel}, {Bonnerot}, {Dipierro}, {Ballabio}, {Ragusa}, {Federrath}, {Iaconi}, {Reichardt}, {Forgan}, {Hutchison}, {Constantino}, {Ayliffe}, {Hirsh}, \& {Lodato}}]{phantom2018}
{Price}, D.~J., {Wurster}, J., {Tricco}, T.~S., {et~al.} 2018{\natexlab{b}}, \pasa, 35, e031, \dodoi{10.1017/pasa.2018.25}

\bibitem[{{Rabago} {et~al.}(2023){Rabago}, {Zhu}, {Martin}, \& {Lubow}}]{rabago2023}
{Rabago}, I., {Zhu}, Z., {Martin}, R.~G., \& {Lubow}, S.~H. 2023, \mnras, 520, 2138, \dodoi{10.1093/mnras/stad242}

\bibitem[{{Rafikov}(2002)}]{rafikov2002}
{Rafikov}, R.~R. 2002, \apj, 569, 997, \dodoi{10.1086/339399}

\bibitem[{{Ragusa} {et~al.}(2017){Ragusa}, {Dipierro}, {Lodato}, {Laibe}, \& {Price}}]{ragusa2017}
{Ragusa}, E., {Dipierro}, G., {Lodato}, G., {Laibe}, G., \& {Price}, D.~J. 2017, \mnras, 464, 1449, \dodoi{10.1093/mnras/stw2456}

\bibitem[{{Ren} {et~al.}(2018){Ren}, {Dong}, {Esposito}, {Pueyo}, {Debes}, {Poteet}, {Choquet}, {Benisty}, {Chiang}, {Grady}, {Hines}, {Schneider}, \& {Soummer}}]{ren2018}
{Ren}, B., {Dong}, R., {Esposito}, T.~M., {et~al.} 2018, \apj, 857, L9, \dodoi{10.3847/2041-8213/aab7f5}

\bibitem[{{Ren} {et~al.}(2020){Ren}, {Dong}, {van Holstein}, {Ruffio}, {Calvin}, {Girard}, {Benisty}, {Boccaletti}, {Esposito}, {Choquet}, {Mawet}, {Pueyo}, {Stolker}, {Chiang}, {Boer}, {Debes}, {Garufi}, {Grady}, {Hines}, {Maire}, {M{\'e}nard}, {Millar-Blanchaer}, {Perrin}, {Poteet}, \& {Schneider}}]{ren2020}
{Ren}, B., {Dong}, R., {van Holstein}, R.~G., {et~al.} 2020, \apjl, 898, L38, \dodoi{10.3847/2041-8213/aba43e}

\bibitem[{{Salyk} {et~al.}(2013){Salyk}, {Herczeg}, {Brown}, {Blake}, {Pontoppidan}, \& {van Dishoeck}}]{salyk2013}
{Salyk}, C., {Herczeg}, G.~J., {Brown}, J.~M., {et~al.} 2013, \apj, 769, 21, \dodoi{10.1088/0004-637X/769/1/21}

\bibitem[{{Speedie} {et~al.}(2024){Speedie}, {Dong}, {Hall}, {Longarini}, {Veronesi}, {Paneque-Carre{\~n}o}, {Lodato}, {Tang}, {Teague}, \& {Hashimoto}}]{speedie2024}
{Speedie}, J., {Dong}, R., {Hall}, C., {et~al.} 2024, \nat, 633, 58, \dodoi{10.1038/s41586-024-07877-0}

\bibitem[{{Speedie} {et~al.}(2025){Speedie}, {Dong}, {Teague}, {Segura-Cox}, {Pineda}, {Calcino}, {Longarini}, {Hall}, {Tang}, {Hashimoto}, {Paneque-Carre{\~n}o}, {Lodato}, \& {Veronesi}}]{speedie2025}
{Speedie}, J., {Dong}, R., {Teague}, R., {et~al.} 2025, \apjl, 981, L30, \dodoi{10.3847/2041-8213/adb7d5}

\bibitem[{{Tang} {et~al.}(2012){Tang}, {Guilloteau}, {Pi{\'e}tu}, {Dutrey}, {Ohashi}, \& {Ho}}]{tang2012-abaur-envelope}
{Tang}, Y.~W., {Guilloteau}, S., {Pi{\'e}tu}, V., {et~al.} 2012, \aap, 547, A84, \dodoi{10.1051/0004-6361/201219414}

\bibitem[{{Tang} {et~al.}(2017){Tang}, {Guilloteau}, {Dutrey}, {Muto}, {Shen}, {Gu}, {Inutsuka}, {Momose}, {Pietu}, {Fukagawa}, {Chapillon}, {Ho}, {di Folco}, {Corder}, {Ohashi}, \& {Hashimoto}}]{tang2017}
{Tang}, Y.-W., {Guilloteau}, S., {Dutrey}, A., {et~al.} 2017, \apj, 840, 32, \dodoi{10.3847/1538-4357/aa6af7}

\bibitem[{Teague \& Foreman-Mackey(2018)}]{bettermoments2018}
Teague, R., \& Foreman-Mackey, D. 2018, {bettermoments: A robust method to measure line centroids}, v1.0,  Zenodo, \dodoi{10.5281/zenodo.1419754}

\bibitem[{{Teague} {et~al.}(2025){Teague}, {Benisty}, {Facchini}, {Fukagawa}, {Pinte}, {Andrews}, {Bae}, {Barraza-Alfaro}, {Cataldi}, {Cuello}, {Curone}, {Czekala}, {Fasano}, {Flock}, {Galloway-Sprietsma}, {Garg}, {Hall}, {Hammond}, {Hilder}, {Huang}, {Ilee}, {Izquierdo}, {Kanagawa}, {Lesur}, {Lodato}, {Longarini}, {Loomis}, {Masset}, {Menard}, {Orihara}, {Price}, {Rosotti}, {Stadler}, {Testi}, {Yen}, {Wafflard-Fernandez}, {Wilner}, {Winter}, {W{\"o}lfer}, {Yoshida}, \& {Zawadzki}}]{exo1}
{Teague}, R., {Benisty}, M., {Facchini}, S., {et~al.} 2025, \apjl, 984, L6, \dodoi{10.3847/2041-8213/adc43b}

\bibitem[{{Terry} {et~al.}(2022){Terry}, {Hall}, {Longarini}, {Lodato}, {Toci}, {Veronesi}, {Paneque-Carre{\~n}o}, \& {Pinte}}]{terry2022}
{Terry}, J.~P., {Hall}, C., {Longarini}, C., {et~al.} 2022, \mnras, 510, 1671, \dodoi{10.1093/mnras/stab3513}

\bibitem[{{Toomre}(1964)}]{toomre1964}
{Toomre}, A. 1964, \apj, 139, 1217, \dodoi{10.1086/147861}

\bibitem[{{Unno} {et~al.}(2022){Unno}, {Hanawa}, \& {Takasao}}]{unno2022}
{Unno}, M., {Hanawa}, T., \& {Takasao}, S. 2022, \apj, 941, 154, \dodoi{10.3847/1538-4357/aca410}

\bibitem[{Uyama {et~al.}(2020)Uyama, Muto, Mawet, Christiaens, Hashimoto, Kudo, Kuzuhara, Ruane, Beichman, Absil, \& et~al.}]{Uyama2020}
Uyama, T., Muto, T., Mawet, D., {et~al.} 2020, The Astronomical Journal, 159, 118, \dodoi{10.3847/1538-3881/ab7006}

\bibitem[{{Valdivia-Mena} {et~al.}(2022){Valdivia-Mena}, {Pineda}, {Segura-Cox}, {Caselli}, {Neri}, {L{\'o}pez-Sepulcre}, {Cunningham}, {Bouscasse}, {Semenov}, {Henning}, {Pi{\'e}tu}, {Chapillon}, {Dutrey}, {Fuente}, {Guilloteau}, {Hsieh}, {Jim{\'e}nez-Serra}, {Marino}, {Maureira}, {Smirnov-Pinchukov}, {Tafalla}, \& {Zhao}}]{valdiviamena2022}
{Valdivia-Mena}, M.~T., {Pineda}, J.~E., {Segura-Cox}, D.~M., {et~al.} 2022, \aap, 667, A12, \dodoi{10.1051/0004-6361/202243310}

\bibitem[{{Veronesi} {et~al.}(2024){Veronesi}, {Longarini}, {Lodato}, {Laibe}, {Hall}, {Facchini}, \& {Testi}}]{Veronesi2024}
{Veronesi}, B., {Longarini}, C., {Lodato}, G., {et~al.} 2024, \aap, 688, A136, \dodoi{10.1051/0004-6361/202348237}

\bibitem[{{Veronesi} {et~al.}(2021){Veronesi}, {Paneque-Carre{\~n}o}, {Lodato}, {Testi}, {P{\'e}rez}, {Bertin}, \& {Hall}}]{Veronesi2021}
{Veronesi}, B., {Paneque-Carre{\~n}o}, T., {Lodato}, G., {et~al.} 2021, \apjl, 914, L27, \dodoi{10.3847/2041-8213/abfe6a}

\bibitem[{{Weingartner} \& {Draine}(2001)}]{weingartner2001}
{Weingartner}, J.~C., \& {Draine}, B.~T. 2001, \apj, 548, 296, \dodoi{10.1086/318651}

\bibitem[{{Wilson} \& {Rood}(1994)}]{wilson1994}
{Wilson}, T.~L., \& {Rood}, R. 1994, \araa, 32, 191, \dodoi{10.1146/annurev.aa.32.090194.001203}

\bibitem[{{Winter} {et~al.}(2024){Winter}, {Benisty}, \& {Andrews}}]{winter2024}
{Winter}, A.~J., {Benisty}, M., \& {Andrews}, S.~M. 2024, \apjl, 972, L9, \dodoi{10.3847/2041-8213/ad6d5d}

\bibitem[{{Winter} {et~al.}(2025){Winter}, {Benisty}, {Izquierdo}, {Lodato}, {Teague}, {Kimmig}, {Andrews}, {Bae}, {Barraza-Alfaro}, {Cuello}, {Curone}, {Czekala}, {Facchini}, {Fasano}, {Hall}, {Hardiman}, {Hilder}, {Ilee}, {Fukagawa}, {Longarini}, {M{\'e}nard}, {Orihara}, {Pinte}, {Price}, {Rosotti}, {Stadler}, {Wilner}, {W{\"o}lfer}, {Yen}, {Yoshida}, \& {Zawadzki}}]{winter2025}
{Winter}, A.~J., {Benisty}, M., {Izquierdo}, A.~F., {et~al.} 2025, \apjl, 990, L10, \dodoi{10.3847/2041-8213/adf113}

\bibitem[{{Woitke} {et~al.}(2019){Woitke}, {Kamp}, {Antonellini}, {Anthonioz}, {Baldovin-Saveedra}, {Carmona}, {Dionatos}, {Dominik}, {Greaves}, {G{\"u}del}, {Ilee}, {Liebhardt}, {Menard}, {Min}, {Pinte}, {Rab}, {Rigon}, {Thi}, {Thureau}, \& {Waters}}]{woitke2019}
{Woitke}, P., {Kamp}, I., {Antonellini}, S., {et~al.} 2019, \pasp, 131, 064301, \dodoi{10.1088/1538-3873/aaf4e5}

\bibitem[{{W{\"o}lfer} {et~al.}(2021){W{\"o}lfer}, {Facchini}, {Kurtovic}, {Teague}, {van Dishoeck}, {Benisty}, {Ercolano}, {Lodato}, {Miotello}, {Rosotti}, {Testi}, \& {Ubeira Gabellini}}]{wolfer2021}
{W{\"o}lfer}, L., {Facchini}, S., {Kurtovic}, N.~T., {et~al.} 2021, \aap, 648, A19, \dodoi{10.1051/0004-6361/202039469}

\bibitem[{{Xie} {et~al.}(2021){Xie}, {Ren}, {Dong}, {Pueyo}, {Ruffio}, {Fang}, {Mawet}, \& {Stolker}}]{xie2021}
{Xie}, C., {Ren}, B., {Dong}, R., {et~al.} 2021, \apjl, 906, L9, \dodoi{10.3847/2041-8213/abd241}

\bibitem[{{Xie} {et~al.}(2023){Xie}, {Ren}, {Dong}, {Choquet}, {Vigan}, {Gonzalez}, {Wagner}, {Fang}, \& {Ubeira-Gabellini}}]{xie2023}
{Xie}, C., {Ren}, B.~B., {Dong}, R., {et~al.} 2023, \aap, 675, L1, \dodoi{10.1051/0004-6361/202346305}

\bibitem[{{Zhang} {et~al.}(2018){Zhang}, {Zhu}, {Huang}, {Guzm{\'a}n}, {Andrews}, {Birnstiel}, {Dullemond}, {Carpenter}, {Isella}, {P{\'e}rez}, {Benisty}, {Wilner}, {Baruteau}, {Bai}, \& {Ricci}}]{zhang2018}
{Zhang}, S., {Zhu}, Z., {Huang}, J., {et~al.} 2018, \apjl, 869, L47, \dodoi{10.3847/2041-8213/aaf744}

\end{thebibliography}
\bibliographystyle{aasjournal}


% \appendix

% \section{Analysis of the spiral arm motion}

\end{CJK*}
\end{document}